\documentclass[preprint,prb,aps,groupedaddress,showpacs,footinbib,amsmath,amssymb]{revtex4-1}
\usepackage{graphicx}
\usepackage{dcolumn}
\usepackage{color}
\usepackage{bm}
\usepackage{natbib,hyperref}
\usepackage{amsmath,amssymb,amsfonts,bbold}
\usepackage[normalem]{ulem}

\newcommand{\beq}{\begin{equation}}
\newcommand{\eeq}{\end{equation}}
\newcommand{\beqa}{\begin{eqnarray}}
\newcommand{\eeqa}{\end{eqnarray}}

\usepackage{verbatim}   
\usepackage{soul}

\begin{document}

\title{Systematics of electronic and magnetic properties in the transition metal
doped Sb$_2$Te$_3$ quantum anomalous Hall platform}
				
\author{M F. Islam}
\author {C.M. Canali}
\affiliation{Department of Physics and Electrical Engineering,\\ Linnaeus
University, 391 82 Kalmar, Sweden}

\author{A. Pertsova} 
\author{A. Balatsky}
\affiliation{Nordita, Roslagstullsbacken 23, SE-106 91 Stockholm, Sweden}

\author{S. K. Mahatha} 
	\affiliation{Istituto di Struttura della Materia (ISM),
Consiglio Nazionale delle Ricerche (CNR), I-34149 Trieste, Italy}
\author{C. Carbone} 
	\affiliation{Istituto di Struttura della Materia (ISM),
Consiglio Nazionale delle Ricerche (CNR), I-34149 Trieste, Italy}
\author{A. Barla} 
	\affiliation{Istituto di Struttura della Materia (ISM),
Consiglio Nazionale delle Ricerche (CNR), I-34149 Trieste, Italy}

\author{K. A. Kokh} 
\author{O. E. Tereshchenko} 
	\affiliation{Novosibirsk State University, 630090 Novosibirsk, Russia}
	
\author{E. Jim\'enez} 
\author{N. B. Brookes} 
	\affiliation{European Synchrotron Radiation Facility, 71 Avenue des Martyrs, Grenoble, France}	
	
\author{P. Gargiani} 
\author{M. Valvidares} 
	\affiliation{ALBA Synchrotron Light Source, E-08290 Cerdanyola del Vall\`es, Spain}	
	
\author{S. Schatz} 
	\affiliation{Physikalisches Institut, Experimentelle Physik VII, 
	Universit\"{a}t W\"{u}rzburg, Am Hubland, D-97074 W\"{u}rzburg, Germany}
\author{T. R. F. Peixoto} 
	\affiliation{Physikalisches Institut, Experimentelle Physik VII, 
	Universit\"{a}t W\"{u}rzburg, Am Hubland, D-97074 W\"{u}rzburg, Germany}
\author{H. Bentmann} 
	\affiliation{Physikalisches Institut, Experimentelle Physik VII, 
	Universit\"{a}t W\"{u}rzburg, Am Hubland, D-97074 W\"{u}rzburg, Germany}
\author{F. Reinert} 
	\affiliation{Physikalisches Institut, Experimentelle Physik VII, 
	Universit\"{a}t W\"{u}rzburg, Am Hubland, D-97074 W\"{u}rzburg, Germany}

\author{J. Jung} 
	\affiliation{Physikalisches Institut, Experimentelle Physik II, 
	Universit\"{a}t W\"{u}rzburg, Am Hubland, D-97074 W\"{u}rzburg, Germany}
\author{T. Bathon} 
	\affiliation{Physikalisches Institut, Experimentelle Physik II, 
	Universit\"{a}t W\"{u}rzburg, Am Hubland, D-97074 W\"{u}rzburg, Germany}	
\author{K. Fauth} 
	\affiliation{Physikalisches Institut, Experimentelle Physik II, 
	Universit\"{a}t W\"{u}rzburg, Am Hubland, D-97074 W\"{u}rzburg, Germany}	
\author{M. Bode} 
	\affiliation{Physikalisches Institut, Experimentelle Physik II, 
	Universit\"{a}t W\"{u}rzburg, Am Hubland, D-97074 W\"{u}rzburg, Germany}
\author{P. Sessi} 
	\affiliation{Physikalisches Institut, Experimentelle Physik II, 
	Universit\"{a}t W\"{u}rzburg, Am Hubland, D-97074 W\"{u}rzburg, Germany}
	
\date{\today}

\begin{abstract} The quantum anomalous Hall effect (QAHE) has recently been
reported to emerge in magnetically-doped topological insulators.
Although its general phenomenology is well established, the microscopic origin is
far from being properly understood and controlled. Here we
report on a detailed and systematic investigation of transition-metal (TM)-doped
Sb$_2$Te$_3$. By combining density functional theory (DFT) calculations with
complementary experimental techniques, i.e., scanning tunneling microscopy
(STM), resonant photoemission (resPES), and  x-ray magnetic circular dichroism
(XMCD), we provide a complete spectroscopic characterization of both electronic
and magnetic properties. Our results reveal that the TM dopants not only affect 
the magnetic state of the host material, but also significantly alter the
electronic structure by generating impurity-derived energy bands.  Our findings
demonstrate the existence of a delicate interplay between electronic and
magnetic properties in TM-doped TIs. In particular, we find that the fate of the topological
surface states critically depends on the specific character of the TM impurity:
while V- and Fe-doped Sb$_2$Te$_3$ display resonant impurity states in
the vicinity of the Dirac point, Cr and Mn impurities leave the 
energy gap unaffected. 
The single-ion magnetic anisotropy energy and easy axis, which control the magnetic gap opening
and its stability, are also found to be strongly TM impurity-dependent and can
vary from in-plane to out-of-plane depending on the impurity and its distance
from the surface. Overall, our results provide general guidelines 
for the realization of a robust QAHE in TM-doped Sb$_2$Te$_3$ in the
ferromagnetic state.
%
\end{abstract}

\maketitle

\section{Introduction} \label{intro} 
Topological insulators (TIs) represent a new state of matter.\cite{hasan2010, XLQi}  
They behave like insulators in the bulk, but are characterized 
by a non-trivial symmetry inversion at the band
gap due to strong spin-orbit coupling. This gives rise to gapless conducting
spin-helical states on their boundaries (edges or surfaces), with linear (Dirac)
energy-momentum dispersion inside the bulk band gap, crossing at the Dirac
point.  The level degeneracy at the Dirac point is topologically protected by
time-reversal symmetry.  

External perturbations that break this symmetry can
open an energy gap at the Dirac point and modify the spin texture, resulting in
several novel quantum phenomena, possibly relevant for spintronic and
magneto-electronic applications.  Among these, one of the most interesting
phenomena is the quantum anomalous Hall effect (QAHE). The idea of a quantum
version of the anomalous Hall effect, i.e., a quantum Hall effect in the absence
of external magnetic field and Landau levels, had already been suggested by Haldane \cite{haldane1988} in
1988 but never realized until recently.  

Soon after the discovery of three-dimensional TIs, it was predicted theoretically that the QAHE should occur
in thin films of magnetic TIs.\cite{Yu02072010}  Key elements leading to the
QAHE are the opening of a magnetic energy gap at the Dirac point of the
topological surface state and the positioning of the Fermi level inside this
gap.  The currently most discussed approach for engineering a gap is magnetic
doping: ferromagnetically ordered magnetic dopants produce a net magnetization
which breaks time-reversal symmetry and, when directed orthogonal to the TI
surface, opens a gap in the TI surface 
states.\cite{LiuPRL09,henk2012topological}  The theoretical prediction of the QAHE was
first verified experimentally\cite{chang2013} in 2013 and since then it has
been confirmed by several other groups (for a recent review, see
Ref.~\onlinecite{Chang_Li2016}).  

However, several outstanding problems remain
and make the QAHE one of the presently most investigated topics of TI 
research.\cite{chang2015, chang2016, peixoto2016, pertsova2016}  For example, the nature
of magnetic order in these systems is not well understood.  
Furthermore, although a precise quantization of the Hall conductivity is achieved,
a small but non-zero and unexplained dissipative longitudinal conductivity remains 
detectable.{\cite{CheckelskyNatPhys2014, BestwickPhys.Rev.Lett.114.187201, chang2015, ChangPhysRevLett.115.057206, chang2016} 
Most importantly, the effect has so far been observed only at ultra-low temperatures
(typically $< 100$\,mK), i.e., well below the Curie temperature of the respective magnetic TI material. 
It is quite self-evident that all these issues are unavoidable side effects of the magnetic impurities, 
which are on one hand necessary to trigger magnetic order but on the other hand also lead to 
unintended changes of the TI's electronic structure near the Dirac point.\cite{sessi2016}

Establishing ferromagnetic order in a TI with an out-of-plane easy axis is a
challenging problem. 
Ideally, it is desirable to attain magnetic TIs with a high
ferromagnetic transition temperature, $T_{\rm C}$, which can be controlled by the
concentration of magnetic impurities.  Since the discovery of TIs, considerable
experimental effort has been dedicated to achieve long-range magnetic order in
two prototypical TIs, Bi$_2$Se$_3$ and Bi$_2$Te$_3$, by doping with different
TM atoms, such as Cr, Mn, or Fe, in both bulk and thin film 
geometries.\cite{bardeleben2013,collins2014,liu2015,figueroa2015,ruzicka2015,scholz2012,li2015,zhang2013,wei2015,mao2013,mao2015}
Significant theoretical work has also been undertaken to understand the mechanism of magnetism in TM doped 
TIs.\cite{abdalla2013,li2013,shelford2012,schlenk2013,canali2014,watson2013,yang2013,
ni2015,schmidt2011,fan2014,eelbo2014,honolka2012,li2012,nunez2012,tran2016,
franz2012} For some of the magnetic TIs where the QAHE has been observed, such
as Cr-doped (Bi,Sb)$_2$Te$_3$, it has been proposed that the mechanism driving
the ferromagnetic transition is based on the Van Vleck spin
susceptibility.\cite{Yu02072010} For an alternative interpretation of the ordering mechanism
in Cr-doped Sb$_2$Te$_3$, see a recent theoretical study published in Ref.~\onlinecite{JKim2017}.  
However, theoretical\cite{Vergniory2014} and experimental\cite{peixoto2016} work has shown that in
V-doped (Bi,Sb)$_2$Te$_3$ (another magnetic TI where the QAHE has been
observed) the exchange mechanism must be more complex than the Van Vleck picture
considered so far, and competing coupling mechanisms involving super-exchange
mediated by $p$-orbitals of the host atoms may be at play.  For example, the presence of a high density of states near the Fermi level and the Dirac point, 
originating from the the 3$d$ levels of the vanadium impurities was recently demonstrated.\cite{peixoto2016}  
These observations are in agreement with the overwhelming majority 
of photoemission and STM studies showing a gapless Dirac cone in magnetically doped 
TIs.\cite{Chen06082010, Xu2012, Valla_PRL2012, scholz12, schlenk2013, Hor_PRB_2010,
okada2011, sessi2014, sessi2016, xu2017}

Yet, the presence of a finite density of states near the Dirac point in magnetic
topological insulators exhibiting the QAHE seems to defy the established basic
theoretical understanding of this effect.  In an attempt to resolve this apparent
contradiction, Sessi {\em et al.},\cite{sessi2016} on the basis of STM data on V-doped
Sb$_2$ Te$_3$ and elaborating on previous theoretical
work,\cite{biswas2010,balatsky2015} proposed a scenario based on the
dual nature of the magnetic TM dopants in a TI, which on one hand are expected
to open a gap at the Dirac point but at the same time provide impurity states in
the same energy region, such that eventually a gapless condition at the
Dirac point is re-established. \cite{sessi2016, zhong2017} The observation of the QAHE is only made possible by the localized nature of the impurity states, which results in the emergence of a mobility gap in samples characterized by a gapless density of states.\cite{sessi2016}

Since the theory supporting this picture was based on a model Hamiltonian that
does not distinguish among different TM impurities,\cite{sessi2016} it is important to carry out
more detailed theoretical and experimental studies assessing the individual
effect that specific TM elements have
on the electronic and magnetic
properties of the respective host TI. This is the scope of the present paper.  Specifically, we
have carried out systematic first-principles studies of the electronic and
magnetic properties of substitutional Cr, V, Fe, and Mn dopants in Sb$_2$Te$_3$
TI, both in bulk and on the surface. In particular, we have computed the
magnetic anisotropy energy and the easy axis for these magnetic TIs. 
In consideration of the relevant role played by the $p$-levels
of the nearby host atoms in the exchange coupling, we have also evaluated the
spin-polarization of the Te and Sb atoms nearest to the TM impurity. Last but
not least, we have focused on the fundamental issue of whether or not different
TM dopants give rise to resonance impurity states in the gap close to the Dirac
point. 

These theoretical predictions will be compared with
experimental findings obtained using complementary techniques.
Scanning tunneling microscopy (STM) has been used to check the structural
properties of the samples while scanning tunneling spectroscopy (STS)
measurements allowed us to evidence the local perturbations induced by the dopants
inside the bulk band gap. Resonant photoemission (resPES) has been used to map
the chemical and orbital character of the bands. Finally, magnetic properties
have been analyzed by x-ray magnetic circular dichroism (XMCD). This technique
allowed us to investigate the emergence of long range magnetic order and to
identify the easy magnetization direction. Even more importantly, it granted
direct access to the magnetic coupling between dopants and the elements of the
host material, shedding light on the exchange mechanisms.

The paper is organized as follows. In Sec.~\ref{Sec:Theory} we describe the results
of our first-principles calculations of the electronic structure
and magnetic properties of TM-doped Sb$_2$Te$_3$, 
providing a theoretical framework to analyze and interpret the experimental studies. 
Sec.~\ref{STM_STS} presents the results of STM and STS measurements, focusing on  
the issue of the impurity resonances appearing in the bulk gap in the vicinity of the Dirac point.
ResPES measurements, providing direct experimental access to the 3$d$ impurity states residing
within the valence band, are discussed in Sec.~\ref{resPES}. 
Sec.~\ref{xmcd} deals with XMCD measurements,
aimed at elucidating the microscopic magnetic characteristics of the TM dopants in the 
TI host, such as the magnetic moment, the anisotropy and the induced spin-polarization on host atoms by the
magnetic dopants. Finally, in Sec.\ref{conclusions} 
we summarize the main findings of our combined theoretical and experimental
study, and discuss their implication for the realization of a robust QAHE regime in ferromagnetic Sb$_2$Te$_3$.   

\section{Theoretical studies} \label{Sec:Theory}

\subsection{Computational details}
\label{computational_sec}
The density functional theory (DFT) calculations are performed using the full-potential all-electron
linearized augmented plane-waves method as implemented in the state-of-the-art
Wien2k {\it ab-initio} package.\cite{wien2k} The Perdew-Burke-Ernzerhof
generalized gradient approximation (PBE-GGA) is used for the exchange correlation functional.\cite{perdew96}
To investigate the trends of electronic properties of different TM impurities in
Sb$_2$Te$_3$ we have constructed bulk and surface supercells (see Fig.\,\ref{struct}) using experimental
lattice parameters ($a = b = 4.264$\,{\AA}, $c = 30.457$\,{\AA},
$\gamma (\angle ab) = 120^\circ$),\cite{anderson1974} with the $c$ axis of the crystal 
(the $z$ axis in our calculations) along the $[001]$ direction. 
For bulk, we have constructed
a $3 \times 3 \times 1$ supercell containing three quintuple layers (QLs) and a total of 135 atoms. 
For surface calculations we have considered a $3 \times 3\times 2$
surface supercell, a slab of finite thickness consisting of six QLs grown along the $[001]$ direction,
which is perpendicular to the (111) surface of Sb$_2$Te$_3$.  
This supercell contains a total of 270 atoms. The slab's bottom and top (111) surfaces are terminated by Te layers.
A vacuum of 15.87\,{\AA}
is added along the surface direction to avoid supercell interaction. 

Impurities are introduced by substituting one Sb
atom per supercell with V, Cr, Mn or Fe, respectively. 
As shown in the left panel of Fig.~\ref{struct}, bulk substitution 
is modeled by replacing an Sb atom in the middle of the supercell,  
corresponding to an impurity concentration of about 11\% for the respective Sb layer 
and 1.8\% for the entire supercell. 
For surface calculations an impurity is added by substituting a Sb atom by a TM atom in the
topmost Sb layer, 
which is the first layer under the very top surface Te layer. 
This doping also corresponds to an impurity concentration in the Sb layers of
11\%.

\begin{figure}[!ht] \centering
\includegraphics[width=0.5\textwidth]{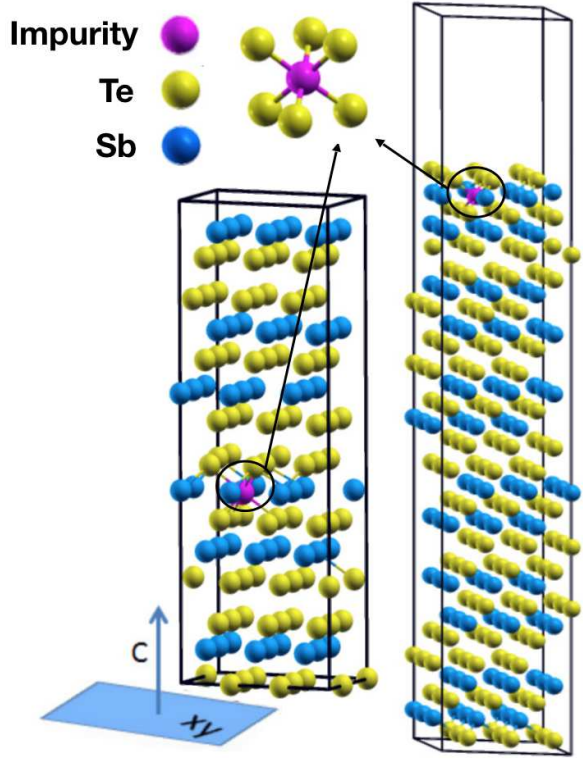} \caption{Bulk (left panel) and surface (right panel)
supercell of Sb$_2$Te$_3$ doped with a TM magnetic impurity substituting a Sb atom of the host
material. The local symmetry of
the impurity atom is $C_{3v}$ for both bulk and surface case. For the surface supercell,
the TM impurity is located on the first Sb monolayer under the top surface consisting of a Te monolayer.
In the surface calculations, the slab thickness consists of six quintuple layers, grown along the $C$ direction. The
exposed top and bottom surfaces of the slab are $(111)$  type.  
} \label{struct}
\end{figure}

All calculations are performed after fully relaxing the atomic positions in all
structures. The effect of relaxation is not very strong in Sb$_2$Te$_3$, both for
bulk and surface. However, the nearest-neighbor Te atoms are observed to move
slightly towards the TM impurity atom. The local symmetry of the impurity atom
is approximately $C_{3v}$ for both bulk and surface except that the distances
between impurity and nearest-neighbor Te atoms are slightly different in bulk
and surface. All the bulk calculations are performed using a $3 \times 3 \times
1$ $k$-mesh, whereas for surface calculations we used $2 \times 2 \times 1$
$k$-mesh in the Brillouin zone.

The single-ion magnetic anisotropy energy (MAE) is calculated by using the force
theorem in the presence of spin-orbit interaction, according to which the
anisotropy energy for two magnetization directions of the TM impurity
is the difference between the sum of all single particle Kohn-Sham eigenvalues
for the two directions.\cite{wang1996}  In this work we have calculated the
energies for two magnetization directions, namely, the out-of-plane [001]
direction (corresponding to the $c$ axis in bulk) and the in-plane [110] direction
(in the $x,y$ plane),
\begin{equation} {\rm MAE} = \sum_{i,[001]}^{E_f}
(\epsilon_{i\uparrow}+\epsilon_{i\downarrow})-\sum_{i,[110]}^{E_f}
(\epsilon_{i\uparrow}+\epsilon_{i\downarrow}). \label{aniso} \end{equation}
Since Wien2k is an all-electron code with plane wave basis, the electronic and
magnetic properties of a system can be calculated very accurately.  The
trade-off is that the calculations carried out with this code is
computationally extremely demanding, particularly for the large supercells considered in
this work, both in terms memory and calculation time.  For example,
a typical surface calculation requires a total of about six weeks to attain a
fully relax crystal structure in the presence of the TM impurity and to perform
anisotropy calculations in a 20 core node with 64 GB memory (which is the
maximum memory/node available in our supercomputer cluster at Lunarc).

Because
of these computational constraints, most of the calculations in this work were
carried out using GGA.  For the case of V-doped Sb$_2$Te$_3$ we have also
performed GGA+U calculations in order to study the effect of electronic
correlation effects which may play a significant role, as indicated by the theoretical 
and experimental data presented below.  DFT in the GGA+U implementation is a suitable
approach to address this case.  For these calculations we have used $U = 4$\,eV,
which is the typical value for TM atoms.\cite{mahadevan2014}

\subsection{Results} \label{result}

\subsubsection{TM impurity magnetic moment and induced spin polarization of the
host atoms}

In pristine Sb$_2$Te$_3$, a semiconductor with bulk gap of $\approx 0.21$\,eV, 
the Sb cations are in the $+3$ oxidation state. 
The 3$d$ TM impurities have an electronic configuration $[{\rm Ar}]3d^n 4s^2$, 
with $n = 3$ for V, $n = 4$ for Cr, $n = 5$ for Mn, and $n = 6$ for Fe.
When a TM atom replaces an Sb atom in Sb$_2$Te$_3$, its possible oxidation states are either
$+3$ or $+2$. In the first case, the TM atom donates the same number of electrons
as the original Sb. 
For this oxidation state, which leads
to the electronic configuration of $[{\rm Ar}]3d^{n-1}$, the expected magnetic moment of the
impurity would naively be $(n - 1) \mu_{\rm B}$. For the oxidation state $+2$, 
the electronic configuration of the impurity is $[{\rm Ar}]3d^n$, resulting in
a magnetic moment that should be close to the atomic value, 
 $n \mu_{\rm B}$, for V, Cr, and Mn, and $(n - 2) \mu_{\rm B}$  for Fe, 
where the minority-spin levels start to be filled. Since in this case
only two electrons of the TM impurity (the two 4$s$ electrons) contribute to binding,  
a hole state is created in the host crystal, formed predominantly by the $p$ states of the Te atoms nearest
to the TM impurity atom with a small admixture of the impurity $d$ states. 
The appearance of hole states localized on the NN anions and in part on the TM impurity is precisely  
what happens in the familiar case of Mn-doped GaAs, where Mn impurities substitute Ga atoms.
This $p$--$d$ hybridization results in a partial spin polarization of the host
$p$ states, which can play an important role in determining the magnetic
properties and the mechanism responsible for ferromagnetism of doped topological insulators.  

Table~\ref{moment} shows calculated magnetic moments of impurity atoms 
(results for both bulk and surface impurities are displayed) together with the induced moment of the
host atoms, along with the character of the magnetic coupling between the
impurity, the nearest-neighbor (NN) Te atoms, and NN Sb atoms (which are the next nearest-neighbor atoms to the TM impurity). 
The impurity magnetic moment for V is close to $2\,\mu_{\rm B}$ for both bulk and surface doping,
suggesting that substitutional V should be in the oxidation state $+3$. The same oxidation state $+3$ characterizes
Cr impurities, for which the magnetic moment is found to be close to $3\,\mu_{\rm B}$.  
In fact, a close inspection of the orbital occupancies for these impurities reveals that the
electron occupation of the majority-spin 3$d$ levels is close to 2.5 electrons for V and 3.4 electrons for Cr.
At the same time, the minority-spin $d$ levels, which are expected  
above the Fermi level, are occupied by approximately 0.5 electrons, as a result of their hybridization
with the $p$ levels of NN Te atoms. In both cases the 4$s$ levels have a very small electron occupation of $\le 0.1$.

Mn impurities are expected to be in the oxidation state $+2$, which is supported by the observation that the 
electron occupancy of the majority-spin 3$d$ levels is 4.5. A net magnetic moment close to 4\,$\mu_{\rm B}$ 
shown in Table I is then the result of these majority electrons and the opposite contribution of $\approx 0.5$ electrons 
occupying the minority-spin $d$ orbitals due to the effect of $p$ level hybridization. 
In contrast with the other impurities, Fe minority states start to be occupied even in the absence of hybridization with host atoms. Indeed, the analysis of the $d$ level occupancy
shows that Fe has 4.6 majority-spin electrons and 1.3 minority-spin electrons. This implies a total of
six electrons occupying the $d$-level core, corresponding to an oxidation state of $+2$. The magnetic moment of Fe is
therefore close to $3.3\,\mu_{\rm B}$.
These considerations on the oxidation state of the TM impurities in Sb$_2$Te$_3$ are in good agreement with
experimental results obtained for Bi$_2$Te$_3$ TI,\cite{TM_BiTe_oxidation2017} 
which is isoelectronic to Sb$_2$Te$_3$.

\begin{table}[t] \caption{ Magnetic moments (in units of $\mu_{\rm B}$) 
for bulk and surface-doped Sb$_2$Te$_3$ for four 
different TM impurities. (The values for the case of surface impurities are in
parenthesis.) The first column displays the magnetic moment of the TM impurity.
The second column is the value of the total magnetic moment induced by the TM impurity on its
nearest-neighbor (NN) Te atoms. 
The +/- sign means that the direction of the induced moment is
parallel/anti-parallel to the impurity $d$ moment, implying a ferromagnetic (FM)/antiferromagnetic (AFM) exchange
coupling (see third column). 
The fourth column displays the exchange coupling (FM or AFM) between the TM impurity and
its NN Sb atoms. Note that the sign of the exchange coupling of is the same for bulk and
surface doping.} \label{moment} \begin{center}
\begin{tabular}{|c|c|c|c|c|} \hline Impurity  &   \multicolumn{2}{c|}{Magnetic
moment ($\mu_{\rm B}$)} & Int. with & Int. with \\ \cline{2-3} type     &
Impurity & Induced moment & NN Te  & NN Sb                               \\ & &
on NN Te       &        &                                     \\ \hline V & 1.98
(2.00)    &   -0.27 (-0.20)   &  AFM   & FM    \\ Cr       & 3.02 (2.90) & -0.33
(-0.24)  &  AFM   & FM    \\ Mn       & 3.97 (4.12) & -0.17 (-0.03)  & AFM   &
FM    \\ Fe       & 3.20 (3.34)     &   +0.33 (0.24)     &  FM    & FM    \\
\hline \end{tabular} \end{center} \end{table}

The impurity magnetic moment induces a spin-polarization on the nearby Te and Sb host atoms,
the strength of which depends on the type of impurity
involved.  In bulk Sb$_2$Te$_3$, the induced moment at the NN Te sites is about
$0.3\,\mu_{\rm B}$ for V, Cr, and Fe, but less than $0.2\,\mu_{\rm B}$ for Mn. On the
surface, the impurity $d$ states are more localized as compared to those in
bulk because of the lower coordination. As a consequence, the $p$--$d$ hybridization is
smaller on the surface, resulting in smaller induced NN moments. Our
calculations also show that the local moments of V, Cr, and Mn couple
antiferromagnetically (AFM) with the induced moment at the NN Te sites, whereas Fe
couples ferromagnetically (FM) with Te in both bulk and surface. However, the
coupling between the impurity atom and the induced moment at the Sb sites is
FM for all four impurities, consistent with experiment (see also discussion below).\cite{sessi2016,mao2015}   

As mentioned above---and this will become more clear 
when we discuss the impurity-induced modification of
local density of states (DOS)---V-doped Sb$_2$Te$_3$ has 
a complex electronic structure with a substantial DOS
at the Fermi level originating from the V $d$-levels.  It is therefore important to investigate how electronic correlations
affect the electronic and magnetic properties of the system.
For this purpose we have investigated V dopants in bulk Sb$_2$Te$_3$ 
within the framework of GGA+U, with $U = 4.0$\,eV. 
The effect of Hubbard $U$ is to localize the $d$ states of the V atom. 
Consequently, the local moment of V increases to $2.6\,\mu_{\rm B}$. 
A closer inspection of the orbital occupancy shows that the enhanced magnetic moment 
is the result of an increase of the majority-spin occupancy from 2.5 to 2.8 and
a simultaneous decrease of the minority-spin occupancy from 0.48 to 0.26. 
Since the hybridization of the $d$ states of the V impurity with its neighboring atoms
decreases with correlations, the induced moments at the NN Te sites decreases to $0.21\,\mu_{\rm B}$.  
However, the magnetic coupling between V and its NN Te sites remains AFM
as in the case of $U = 0$. 

\subsubsection{Magnetic anisotropy energy of TM impurities in Sb$_2$Te$_3$ }

According to a minimal two-dimensional continuum model,
to open a gap at the DP of the surface states in a magnetic TI, 
it is necessary that the magnetization be oriented along the normal of the TI surface\cite{Zyuzin2011a, Zyuzin2011b, Taskin2017}.  
The orientation of the magnetization is determined by the magnetic anisotropy energy (MAE) of the system. 
We have calculated both the bulk and surface MAE and the
corresponding easy axes for all impurities studied in this work.  
The results are shown in Table~\ref{anisotropy}.
\begin{table}[t] \caption{Single-ion magnetic anisotropy energy (MAE) of
different TM magnetic impurities in bulk and surface Sb$_2$Te$_3$topological
insulator. The out-of-plane easy axis in bulk is along the growth $c$ direction of
the crystal. 
For surface impurities, the out-of-plane easy axis is
along the $[001]$ direction, which is perpendicular to 
the exposed $(111)$ surface, terminating the slab.} \label{anisotropy} \begin{center} \begin{tabular}{|c|c|c|c|c|}
\hline Impurity &   \multicolumn{2}{c|}{Bulk}  & \multicolumn{2}{c|}{Surface} \\
\cline{2-5} type       &  MAE   & Easy axis             & MAE   & Easy axis \\ &
(meV)  &                             & (meV) &                     \\ \hline V
& 0.07    & c axis                   & 0.45    & In-plane \\ Cr       & 0.43
& c axis                   & 0.41    & Out-of-plane \\ Mn & 0.58     & c axis
& 0.65   & In-plane          \\ Fe       & 0.20  & xy-plane                 &
0.60    &  in-plane        \\ \hline \end{tabular} \end{center} \end{table}
The direction of easy axis depends on the type of impurity. For V, Cr, and Mn
bulk impurities, the easy axis is out-of-plane, that is, along the the
growth direction of the crystal  cleavage plane, while for Fe it is within the surface plane. Our results show that the surface anisotropy may qualitatively differ from the 
corresponding bulk values. For example, V and Mn exhibit an easy axes within the surface plane, 
contrary to bulk, whereas the Cr easy axis is still out-of-plane, just like for bulk.
These calculations suggest that only the ground
state of Cr-doped Sb$_2$Te$_3$ can sustain a ferromagnetic order with out-of-plane
magnetization, and therefore will open a magnetic gap at the DP. 

Whereas the calculated out-of-plane easy axis of Cr-doped Sb$_2$Te$_3$ 
agrees well with experiment,\cite{duffy2017} theoretical data for V-doped samples 
suggest an in-plane anisotropy for the surface layer which is not supported by the experiment 
(see also the discussion in Sec.~\ref{xmcd}).\cite{chang2015} 
To understand this apparent discrepancy between the experimental measurement and
the theoretical results we would like to emphasize that in our DFT calculations
the TM impurity has been added only to the topmost Sb layer (i.e., the second
sub-surface layer). Therefore in our surface calculations we have not really
considered any magnetic order originating from bulk doping.  If bulk magnetic order is also achieved, it can act as an
external magnetic field to the surface electrons and can significantly influence
the anisotropic properties of the film.  Thus, it might be possible that bulk
magnetic order can change the orientation of magnetization produced by a single
surface layer.

\subsubsection{Resonances in bulk-doped Sb$_2$Te$_3$} \label{Subsec:ImpRes}

\begin{figure}[t] \centering
\includegraphics[width=7.5cm,height=4.5cm]{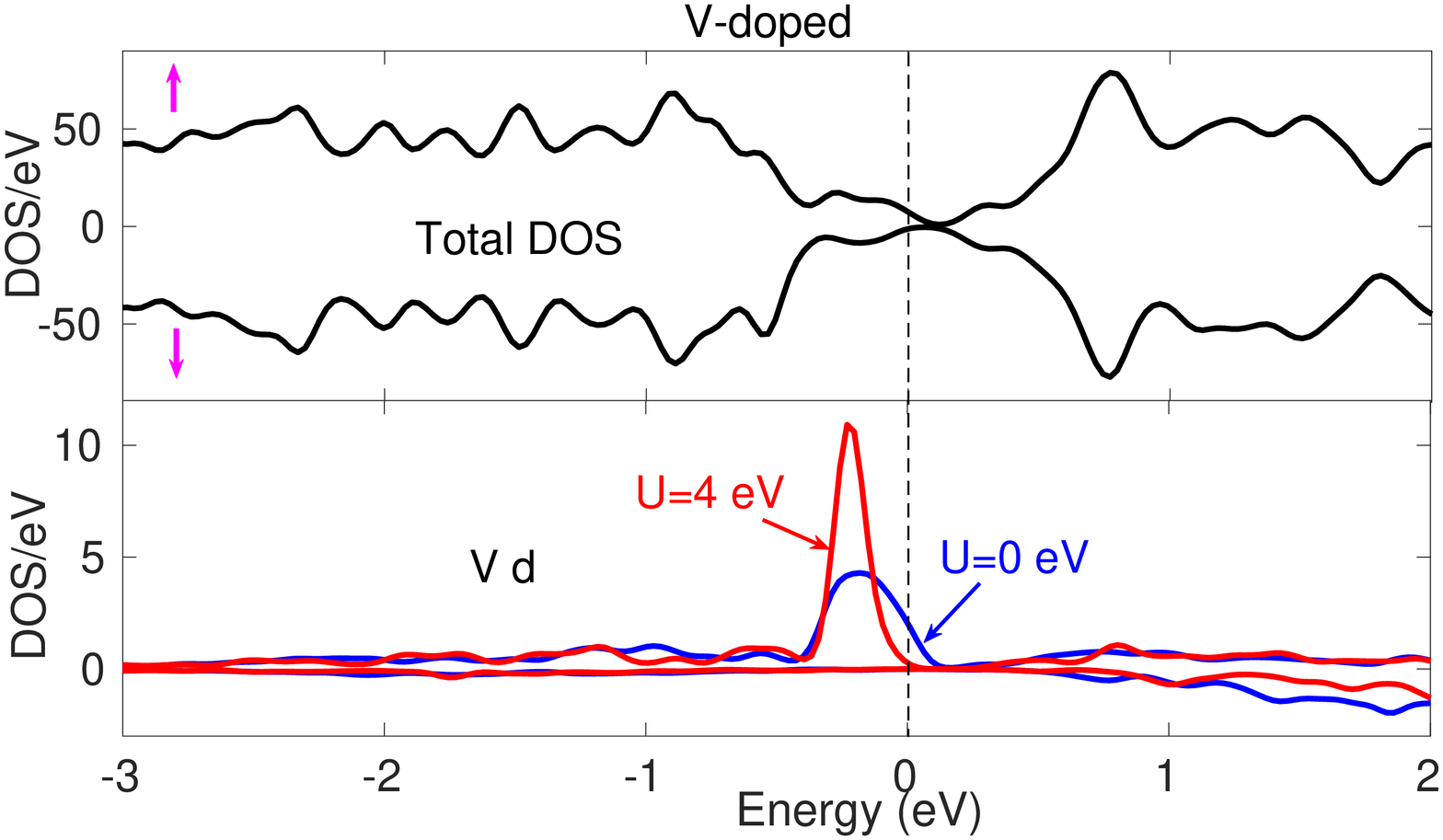} 
\includegraphics[width=7.5cm,height=4.5cm]{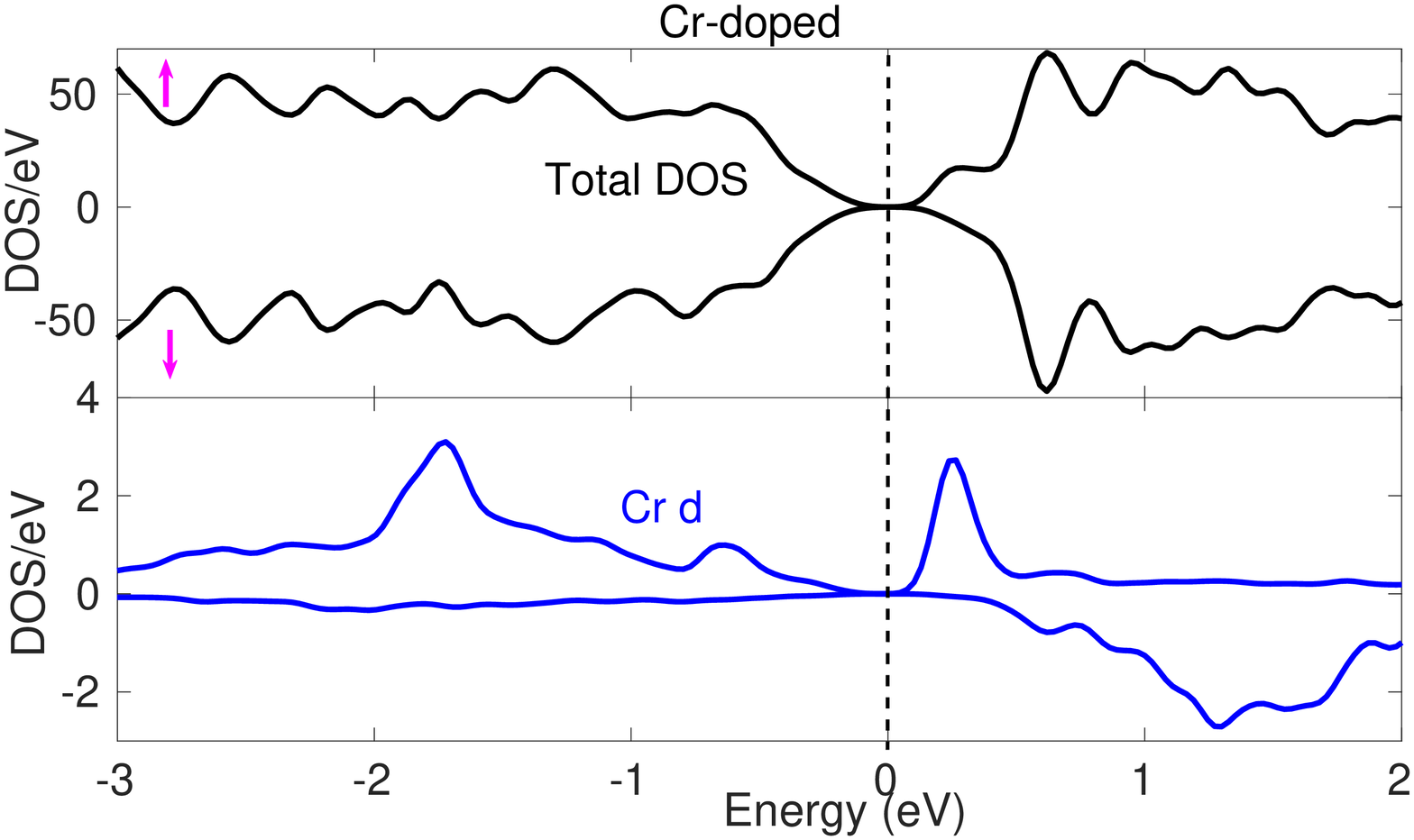} 
\includegraphics[width=7.5cm,height=4.5cm]{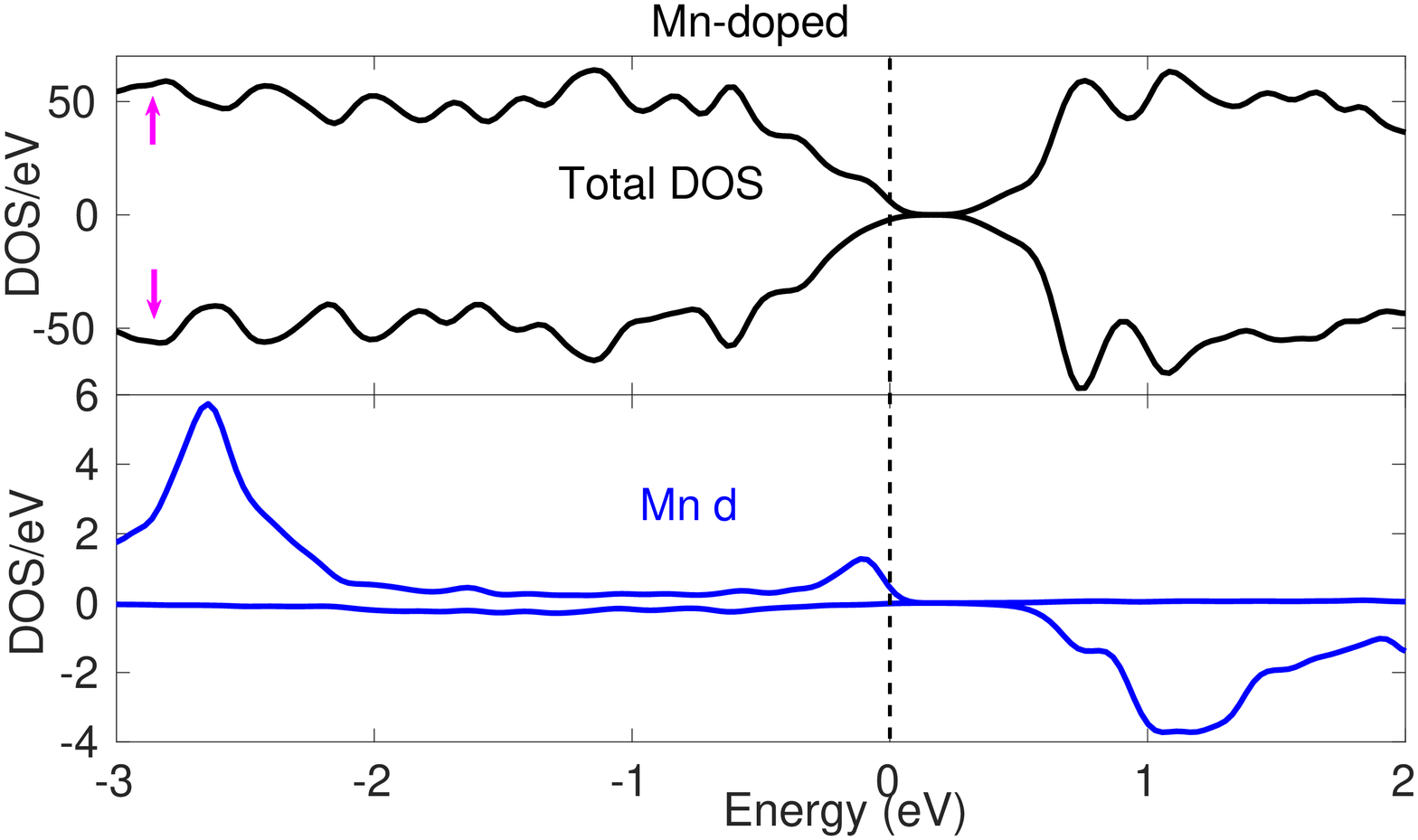} 
\includegraphics[width=7.5cm,height=4.5cm]{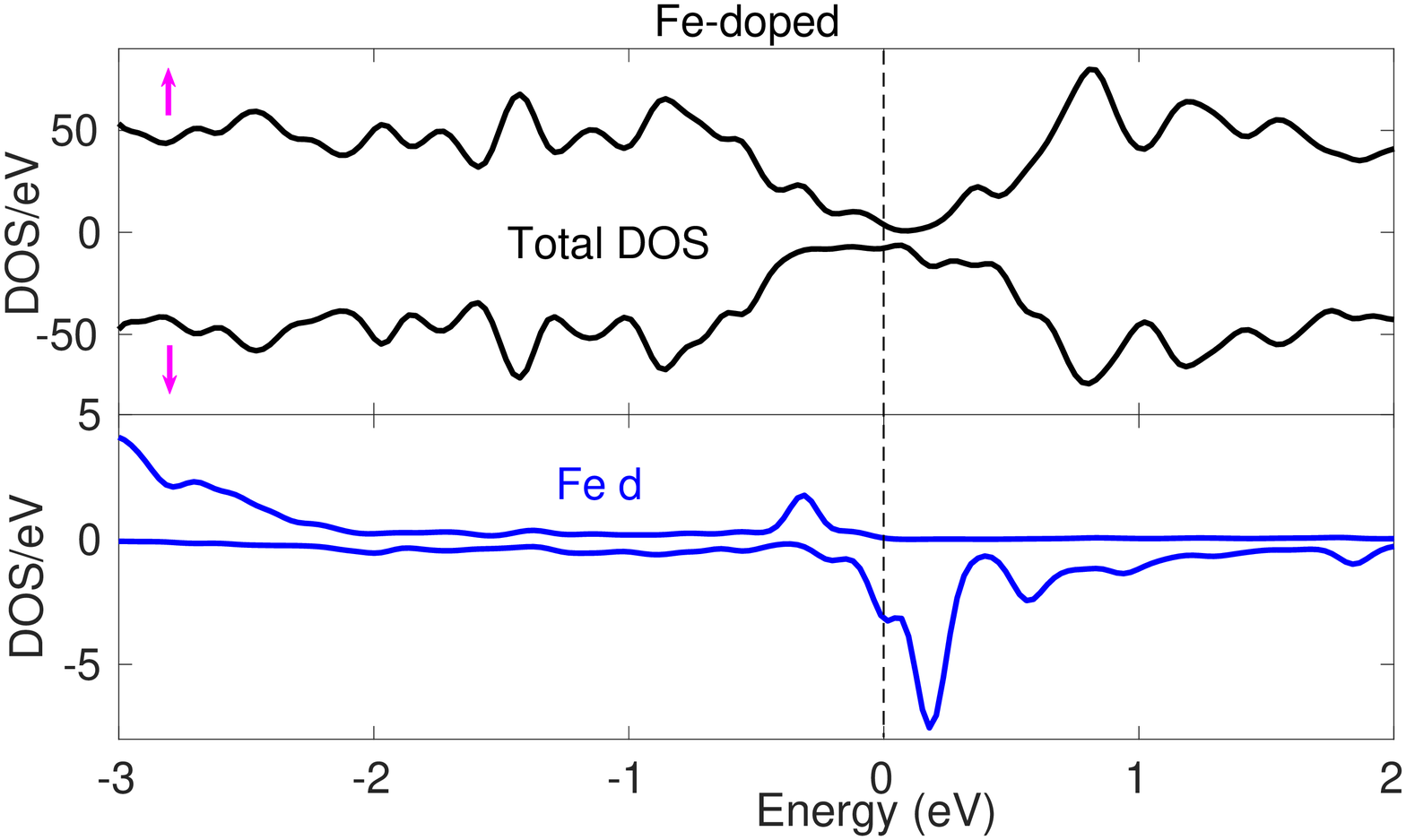}  \caption{Density of
states of bulk TM-doped Sb$_2$Te$_3$ for different bulk magnetic impurities. 
Top to bottom: V, Cr, Mn, Fe. For each impurity case, the black curve in the upper panel is the
total DOS, while the blue curve in the lower panel represents the $d$-level DOS of the
impurity. Positive and negative DOS values refer to majority- and 
minority-spin DOS, respectively. The vertical dotted line at zero
energy marks the position of the Fermi level. The red curve for the
V-impurity is the $d$-level DOS obtained in a GGA$+U$ calculations, with $U= 4$ eV.
All other results are for $U= 0$.}
\label{BulkDOS} \end{figure}
As discussed in Sec.~\ref{intro}, the presence of impurities can give rise to
impurity resonances close to the DP, which can be detrimental on the observation of the QAHE.
To investigate whether the TM impurities give rise to resonance states, we have
first calculated the {\it bulk} density of states (DOS) for TM-doped systems. 
In particular, we are interested in determining the position of
the impurity states in the bulk gap relative to the Fermi level. The total
density (top panel in each of the four sub-figures) and the density of $d$ states 
(bottom panel in each of the four sub-figure) of various bulk dopants is plotted in Fig.~\ref{BulkDOS}.  
The most important feature shown in the figure is that for both V and Mn dopants there is a
finite density of impurity states at the Fermi level in the form of a broad resonance state
(the Fermi energy is the vertical dotted line in the figure).  
On the other hand, for Cr the Fermi level is just
in the middle of the gap, which appears to remain free of resonance states. Impurity $d$-level states for Cr 
(see lower panel for the Cr case) all occur inside the bulk conduction band.  
Fe dopants are qualitatively similar to V, displaying a finite DOS at $E_{\rm F}$
originating from the minority-spin $d$-levels. In contrast, Mn is qualitatively
more similar to Cr, displaying only a small DOS at $E_{\rm F}$, which however is
located not in the middle of bulk gap as for Cr, but closer to the valence band edge.

Given the presence of impurity states in the gap for V dopants, it is important
to assess the role played by electron correlations, which can modify
the electronic structure of the impurity states. Therefore, we have carried out GGA+U calculations 
with $U = 4$\,eV for V-doped Sb$_2$Te$_3$ .  The resulting V $d$-level DOS calculated 
is shown as a red line in Fig.~\ref{BulkDOS} (top panel).
As expected, the effect of the on-site Hubbard interaction $U$ is to increase the localization of the
$d$-electrons around the impurity core. 
This is reflected in the $d$-level DOS of the majority-spins, which are pushed further below the Fermi level, 
increasing their overall occupancy, as we have already seen above in the discussion of the magnetic moment.
The resonant nature of these impurity states is enhanced by the correlations:  the broad shoulder at the Fermi energy becomes a higher and narrower peak. 
However, it is important to note that, although correlations have the effect of reducing the $d$-level DOS at the
Fermi energy, the DOS near $E_{\rm F}$ is still finite.  Although not calculated explicitly here, previous
calculations on TM-doped Bi$_2$Se$_3$ show that the inclusion of
Hubbard $U$ has a very similar effect on the $d$-level DOS of other TM
impurities.\cite{canali2014}  In particular, we expect that correlations will produce minimal
changes to the Cr $d$-level DOS, and by displacing the $d$-levels in the way
described above, will make the Mn DOS look even more similar to Cr. 
The case of Fe is a bit peculiar, since the finite DOS in the absence of correlations is due
to minority-spin $d$-levels.  However it is likely that the effect of $U$, while
pushing these levels further above $E_{\rm F}$, will nevertheless maintain a
finite, albeit small, DOS coming from these states.  In conclusion, our calculations show that only 
for V and---to a lesser extent---Fe dopants resonant states are theoretically expected inside the bulk gap close to $E_{\rm F}$. The DOS of these $d$-levels remains
finite even when electron correlations are included.

\subsubsection{Resonances in surface-doped Sb$_2$Te$_3$} \label{Subsec:SurfRes}

\begin{figure}[b] \centering 
\includegraphics[width=8cm,height=3.5cm]{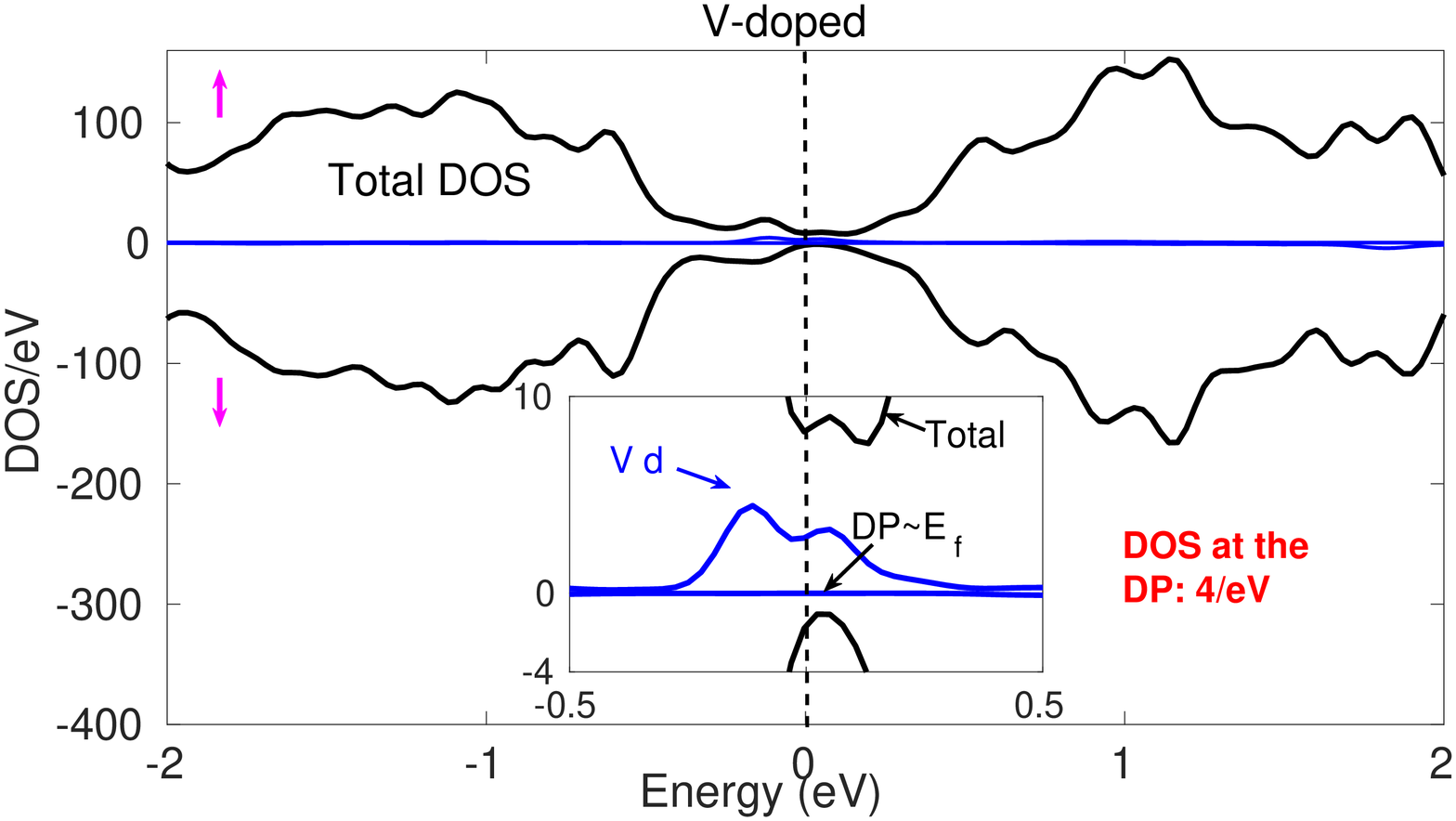}
\includegraphics[width=8cm,height=3.5cm]{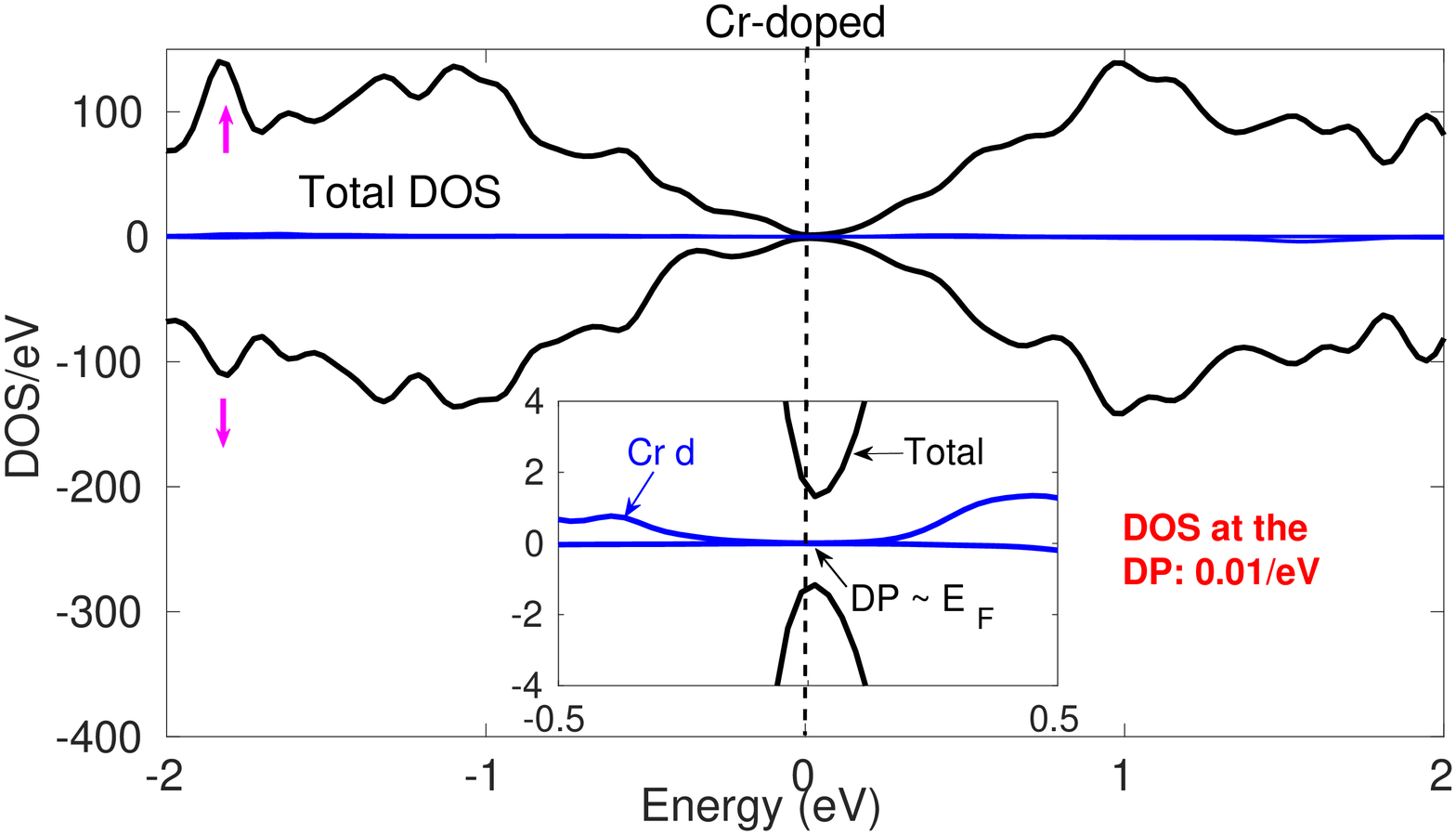}
\includegraphics[width=8cm,height=3.5cm]{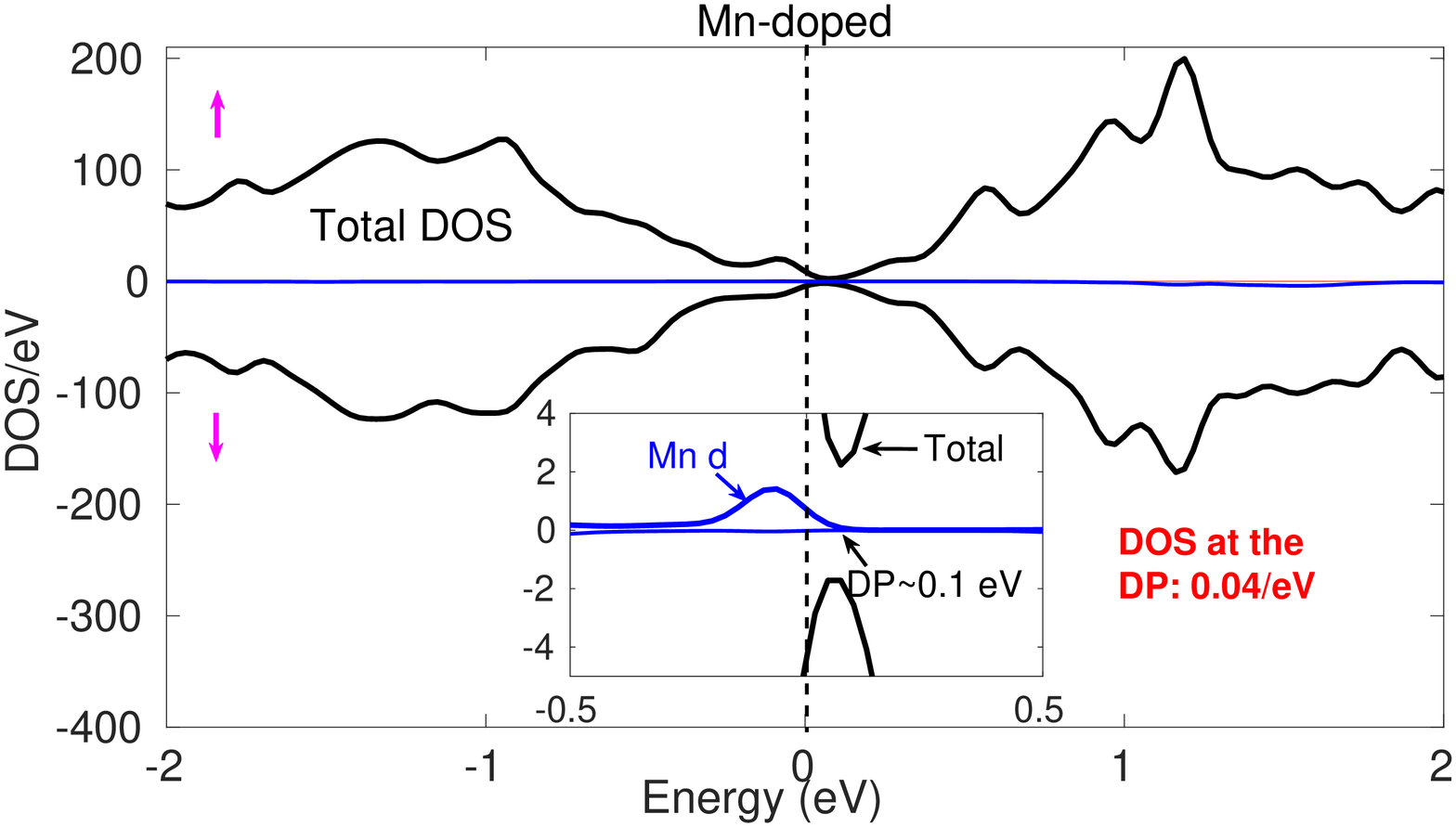}
\includegraphics[width=8cm,height=3.5cm]{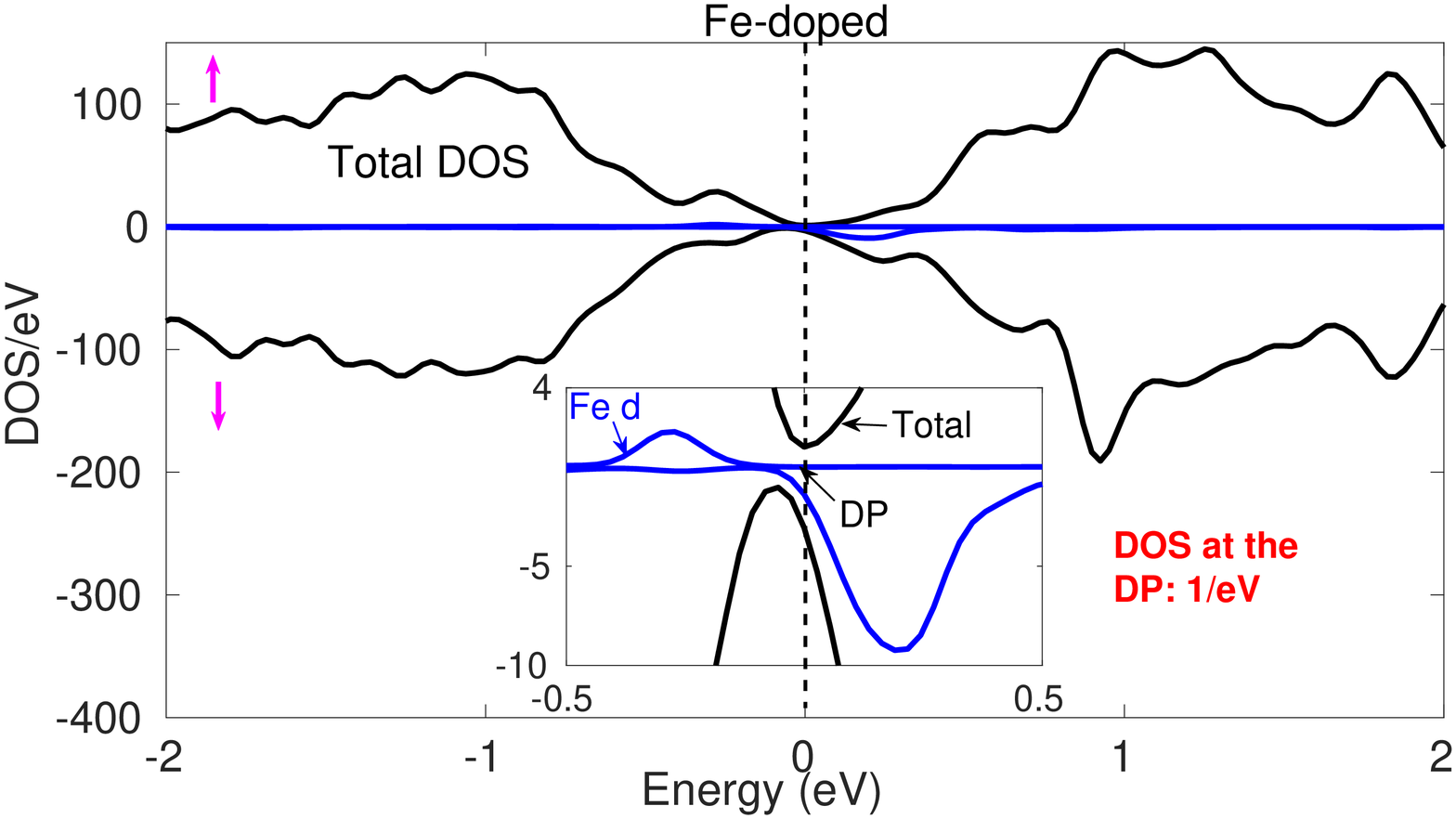} \caption{Density of
states of TM-doped Sb$_2$Te$_3$ for different surface magnetic impurities.  
Top to bottom: V, Cr, Mn and Fe. 
The vertical dotted line at zero
energy marks the position of the Fermi level.
The inset in each plot
shows the DOS around the Dirac point (DP) of the top-surface states. The blue curve in this
inset is the impurity $d$-level DOS. 
For all dopants the DP is very close to the
Fermi level, with the exception of Mn, where the DP is $\approx $ 0.1 eV.
The value of 
the impurity $d$-level DOS at the DP is indicated in red. 
For V there is a finite $d$-level DOS at the DP. On the other hand, for Cr the $d$-level DOS at the DP 
is essentially zero. Fe and Mn are somewhat in between these two extreme, 
with Fe resembling V and Mn resembling Cr.}
\label{SurfDOS} \end{figure}  

We now consider the case of TM magnetic impurities located on the (111) surface of
Sb$_2$Te$_3$, focusing on their effect on the electronic DOS in the vicinity of
the Dirac point. As  explained in Sec.~\ref{computational_sec}, the phrase 
{\it impurity located on the surface} is used to describe a TM atom replaces an Sb atom 
in the topmost Sb layer, just underneath the Te layer terminating the supercell 
(see Fig.~\ref{struct}). This TM atom primarily hybridizes with the three NN Te atoms in the topmost layer 
and another three in the third layer.
The penetration of the topological surface states inside the bulk material 
decays exponentially with the distance from the surface.
More precisely, it is known that the decay length of the
surface states is of the order of two to three quintuple layers (QLs).\cite{NJP}
Therefore we expect that the orbital levels of the TM impurity will overlap
with and strongly affect the surface states of the top surface of the supercell.
On the other hand the TM impurity will have essentially no influence on the
surface states of the bottom surface of the supercell, which is six QLs thick.
The theoretical analysis of the TM impurities at the surface considered here is
essential for the interpretation of the STM and XMCD experiments
described below, which are mostly surface sensitive.


The DOS results of the surface impurity calculations are plotted in Fig.~\ref{SurfDOS}. 
Comparing with Fig.~\ref{BulkDOS}, it is evident that the impurity states for 
surface doping resemble in part the DOS for bulk doping. 
In order to identify the precise position of the DP relative to the Fermi level, the band structure
is required. However, since we were unable to perform band structure calculations for such a large
supercell, due to the computational limitations explained in Sec.~\ref{computational_sec}, 
the DP is estimated by assuming that the DOS is expected to be a minimum at the DP, as 
shown in Fig.~\ref{SurfDOS}. We would like to emphasize that the presence of a finite DOS around the DP
of the top-surface states is due to the presence of the bottom-surface states at the same energies. 
Indeed, as shown in Ref.~\onlinecite{canali2014}, when an impurity is positioned only on the top surface of a
finite-thickness slab, inversion symmetry is broken, and the two Dirac cones of 
top and bottom surfaces are shifted with respect to each other. 
As a result, the energy region containing the DP of the top surface states 
displays a finite DOS arising from the shifted surface states of the bottom surface.

For V-doped Sb$_2$Te$_3$, the impurity $d$-states clearly lie not only inside the
bulk gap, but also quite close the DP of the surface states, in the energy region that
should precisely correspond to a magnetic gap. This important theoretical
result is consistent with experiment.\cite{sessi2016} 
(Note that in our calculations the DP for V-doped Sb$_2$Te$_3$ 
essentially coincides with the Fermi level.)
For Cr-doped Sb$_2$Te$_3$, the situation is quite different. 
As shown in the second panel of Fig.~\ref{SurfDOS}, in this case there
are no impurity resonance states present within the bulk gap. The Fermi level lies just in
the middle of this bulk gap, and it is very close to the DP.
Thus for this dopant, when the broken time-reversal symmetry of the magnetic state
opens a magnetic gap at the DP, there will be no impurity states available to fill this gap. 
Most likely this is one of the main reasons
why Cr-doped (Bi,Sb)$_2$Te$_3$, a TI with properties similar to Sb$_2$Te$_3$,
remains one of the most suitable systems to realize the QAHE.\cite{chang2013}

The case of Mn surface-doping is displayed in the third panel of Fig.~\ref{SurfDOS}.
Here we can see that Mn impurities introduce a small albeit finite DOS at the Fermi level, which, 
as for the case of bulk impurities, is located close to the valence band edge. 
However, we find that the DP is located approximately
at 0.1\,eV above the Fermi level, where the contribution of Mn impurity states is essentially zero.
Since the magnetic gap possibly opened by the time-reversal breaking perturbation 
is centered at the original DP of Sb$_2$Te$_3$,
and is expected to be of the order of a couple of tens of meV at most, we conclude that in this case the Mn impurity states
should not be able to fill this gap. 
Therefore Mn-doped Sb$_2$Te$_3$ is in a way qualitatively similar 
to Cr-doped Sb$_2$Te$_3$, in that any magnetic gap opened by the impurity will not be affected by
the presence of additional impurity resonances. Notice also that, on the basis of our 
analysis on the effect of correlations discussed in
Fig~\ref{BulkDOS}, we expect that the inclusion of on-site correlation effects
via a Hubbard interaction $U$ will  further remove the
small DOS from the gap region.  

Finally the case of Fe surface-impurities is discussed in the bottom panel of Fig.~\ref{SurfDOS}.
As for the case of bulk impurities, Fe is a bit different from the other three cases, since
the impurity-induced resonant levels inside the bulk gap now originate from minority-spin states. 
In this case we find that the DP, which in our calculation occurs very close to the Fermi level, 
is located in a region where the $d$-level DOS is small but definitely finite, similarly to what happens with V.
However, in contrast to the case of V, where the impurity states merge with the valence band edge, 
these states are now located in the middle of the bulk gap, and therefore resemble more properly an impurity band.


The different electronic behavior of the TM impurities inside the
gap, already visible from the TM $d$-level impurity states, is also reflected in
their coupling to the $p$-levels of the NN Te atoms. 
In Fig.~\ref{SurfTeDOS} we plot the DOS of the impurity $d$-states, together with
the DOS of the $p$ states of its NN Te atoms on the top surface. 
The position of the top-surface state's DP is indicated in the figure by a dashed vertical line.
For comparison, we also plot the DOS 
of the $p$ states of the Te atoms on the bottom surface, which are the NN of a Sb atom placed 
in the equivalent position of the one substituted with the impurity on the top surface.
Since the impurity atoms are placed only in the topmost Sb layer (second layer from the top), only the top DP
is influenced by the impurity, whereas the bottom DP remains unchanged. 

\begin{figure}[t] \centering
\includegraphics[width=8cm,height=3.5cm]{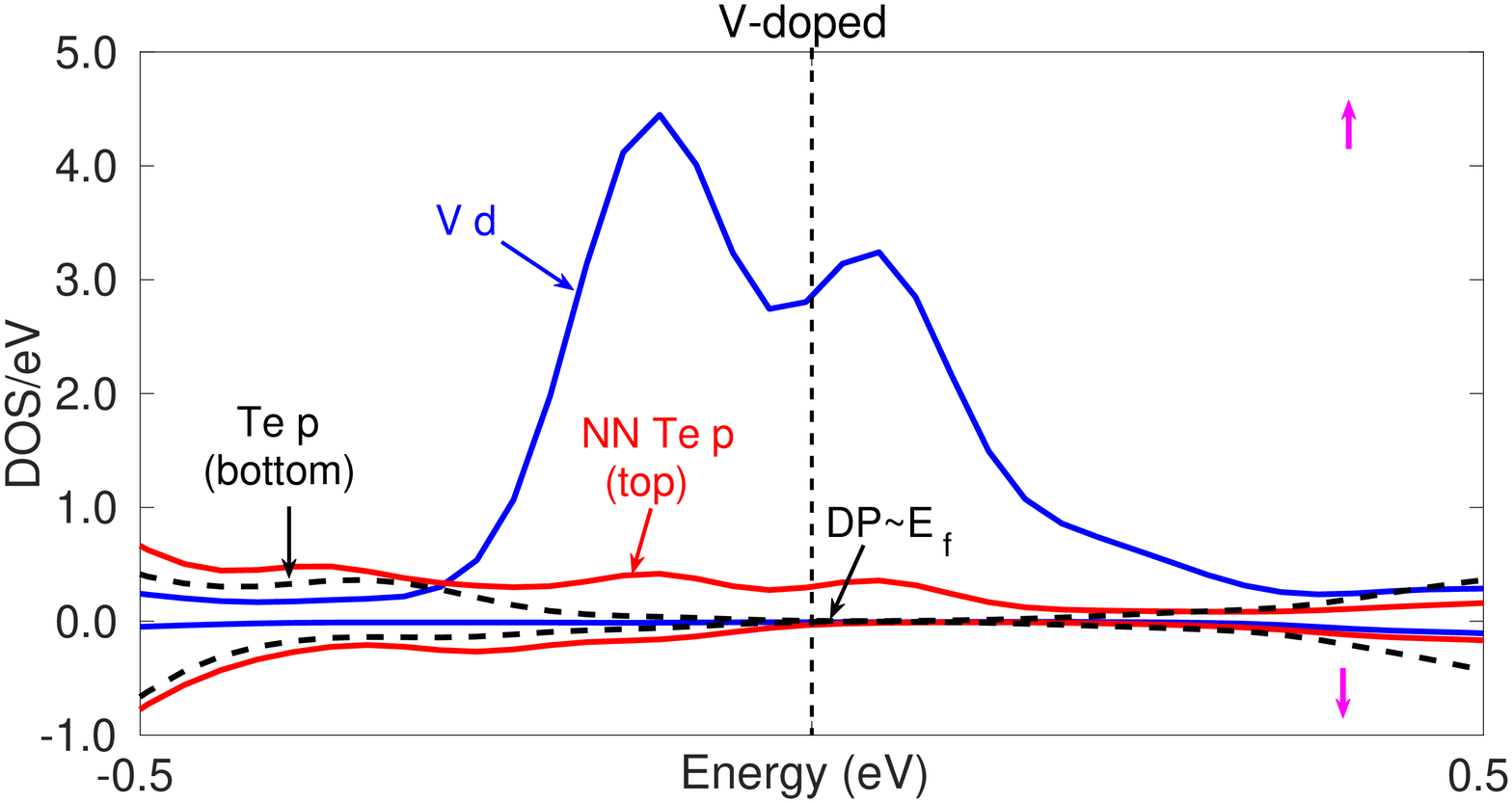}
\includegraphics[width=8cm,height=3.5cm]{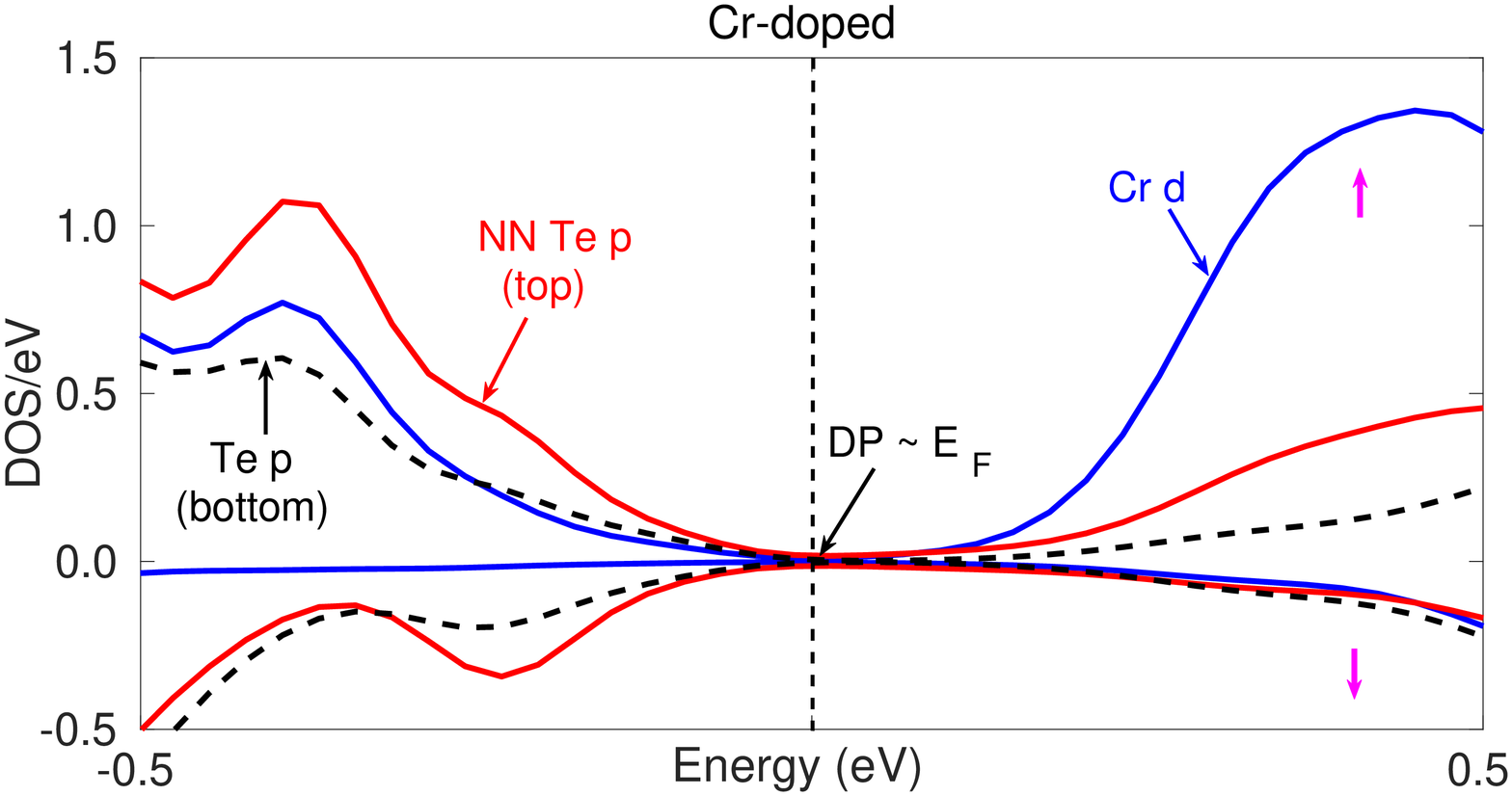}
\includegraphics[width=8cm,height=3.5cm]{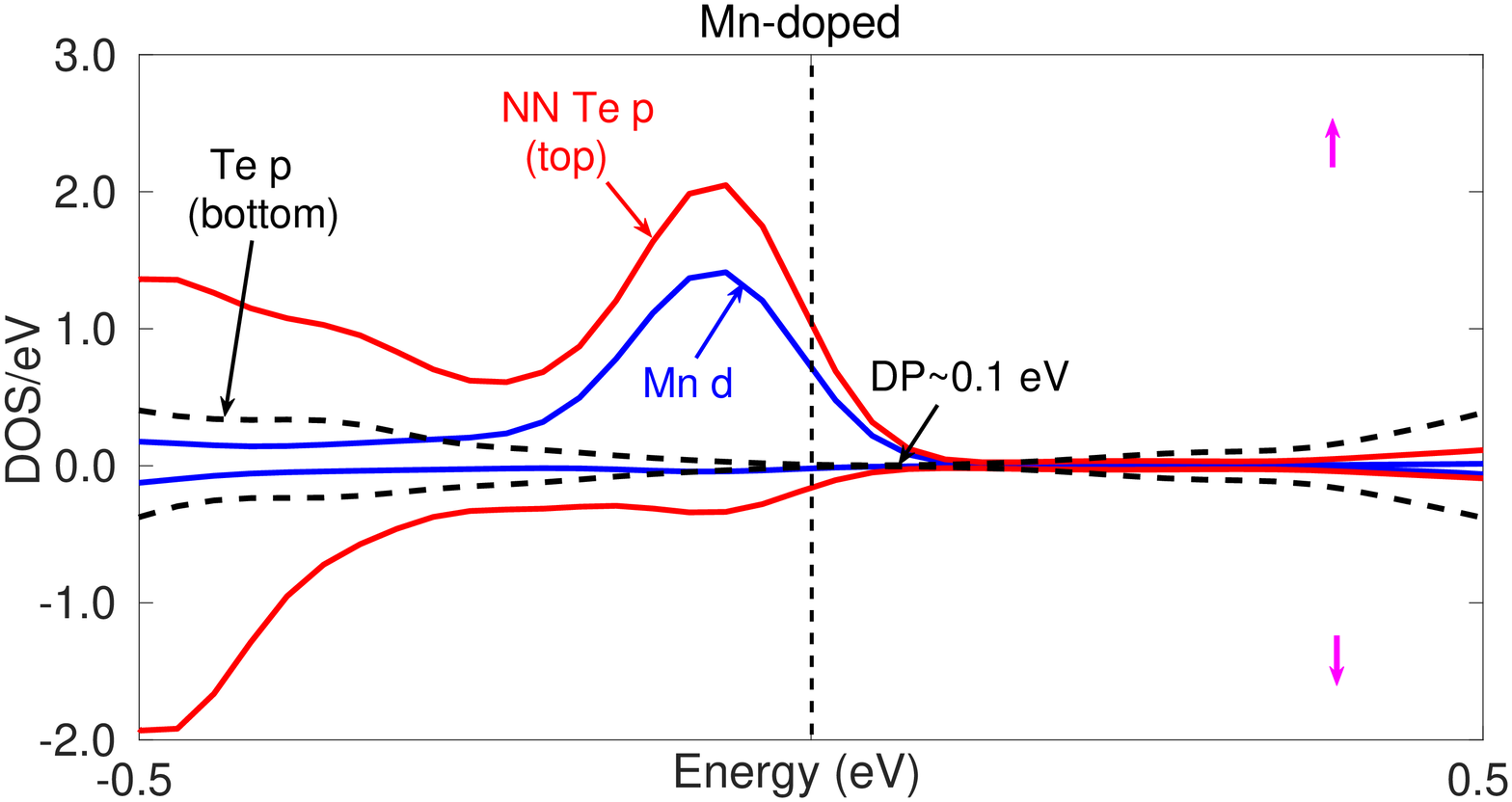}
\includegraphics[width=8cm,height=3.5cm]{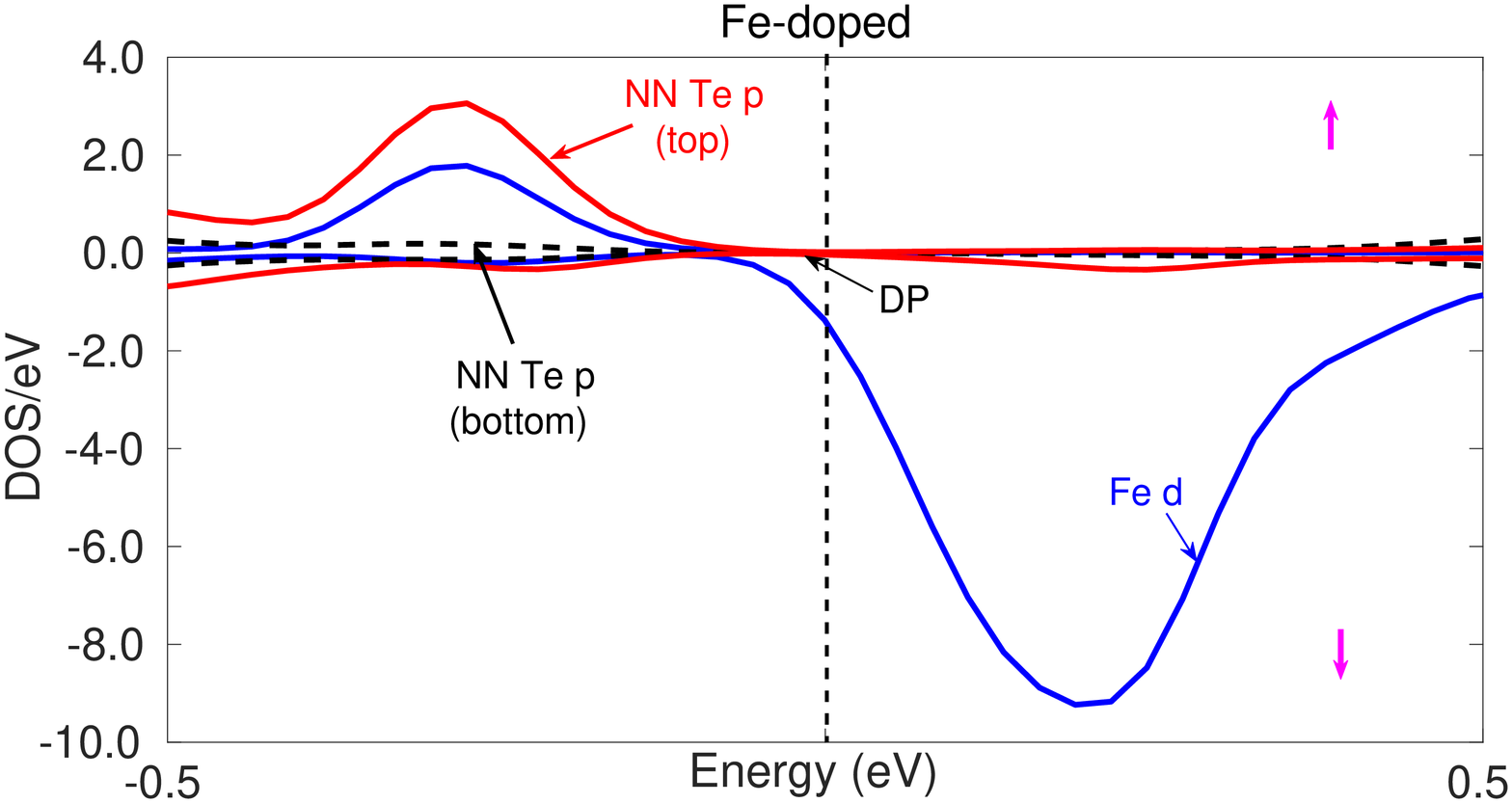} 
\caption{Relative contribution of the impurity $d$-level DOS (blue curve) 
and the NN Te $p$-level DOS of the top
surface (red curve) 
of Sb$_2$Te$_3$ for
different surface magnetic doping. The Te $p$-level DOS of the top surface 
is strongly affected by the nearby impurity. In contrast, the Te $p$-level DOS of the bottom surface,  
also plotted as a comparison (black dashed line) is not influenced by the impurity.} \label{SurfTeDOS} \end{figure}
We note that while the contribution of $p$ states of Te atoms at the
bottom surface is very similar for all types of impurities, at the top surface
the DOS of NN Te $p$ states typically follows closely the impurity $d$ states
due to the $p$--$d$ hybridization (with the exception of Fe), and therefore is quite different for
different impurities. In particular, the striking difference between
V and Cr, already observed in Fig.~\ref{SurfDOS}, is confirmed by these results. For V, there is a finite DOS
of host $p$ states around the Fermi level and the DP, which follows the same trend of the impurity $d$-levels, and extends
above the Fermi level. On the other hand, for Cr this $p$-level contribution from the NN Te atoms is totally absent at the DP. 
For Mn, the $p$-level DOS of the NN Te atoms follows closely the impurity $d$-level DOS. In particular, it is
finite at the Fermi level, but negligible at the DP, which is about 0.1\,eV
above the Fermi level. Note that the behavior of the hybridized NN $p$-levels, 
clearly giving rise to a hole state around the Fermi level,
is similar to what happens for substitutional Mn impurities in GaAs.\cite{fi_cmc_prb_2012}

Fe is again quite different from all other impurities, also concerning the $p$--$d$ hybridization with the NN  atoms.  
The bottom right panel of Fig.~\ref{SurfTeDOS} shows that the
 finite DOS of minority-spin $d$-level impurity states in the vicinity of the DP 
does not result in any appreciable $p$-level contribution from the NN Te atoms. 
Substantial $p$--$d$ hybridization takes place only with majority-spin states below the Fermi level, in an energy region
far away from the DP. Therefore for Fe, the impurity levels emerging in the crucial energy region around the DP, where 
a magnetic gap is likely to open, have mainly $d$-character. These states tend to form an impurity band inside the
bulk band gap, separated from the valence band top edge, and eventually merge with the conduction band bottom edge.

In summary, our calculations show that Mn- and Cr-doped Sb$_2$Te$_3$, in
contrast with V-doped Sb$_2$Te$_3$, do not host any resonance states in the energy region around the DP, where a
magnetic gap in the surface states can be opened by the magnetic impurities.
Our calculations also show a finite DOS of impurity levels at the DP for Fe-doped
Sb$_2$Te$_3$; but, unlike V, these minority-spin states have essentially a $d$-character, and do not hybridize with the
$p$-levels of the NN Te atoms of the host material.  

We conclude this section with a few considerations on the possible nature of the exchange coupling and 
the ensuing mechanism of magnetic order in Sb$_2$Te$_3$ for different dopant species.
The presence of a finite density of impurity $d$ and
host $p$ states around the Fermi level for V and the absence of these states for
Cr, suggests that the mechanism of magnetism is likely to be quite different for V-doped and Cr-doped
Sb$_2$Te$_3$.  In the case of V the impurity states tend to form a rather broad band of resonant states
filling most of the bulk gap and merging with the top of the valence band. This seems to point either
to a RKKY or to a super-exchange interaction mediated by $p$-carriers 
as the most likely source of exchange coupling among the 
magnetic impurities\cite{Vergniory2014}. 
Cr-doped Sb$_2$Te$_3$ is clearly quite different, since the impurities
do not seem to introduce free carriers. Therefore, this case appears to be the best candidate for the
Van Vleck mechanism of magnetic order, as originally suggested in Ref.~\onlinecite{Yu02072010}. However, this 
conclusion has been challenged by a recent paper, reporting the results of first-principles
calculations in Cr-doped Sb$_2$Te$_3$.\cite{JKim2017} This study indicates that magnetic order is present
even in the absence of spin-orbit interaction, which is the driving mechanism
of band inversion and a non-trivial gap in TIs, ultimately responsible for establishing a strong Van Vleck spin susceptibility. 
According to this study, 
the mechanism responsible for 
magnetic order is instead due 
to the presence of an induced FM spin-polarization  in the Sb monolayer containing the
Cr-dopant, transferred widely from the magnetic TM impurity to distant Sb atoms
by the long-range
resonant $p$-bonding network known to exist within the QLs of the chalcogenide host material.
Although we have not carried out any specific study of the  Cr-Cr exchange coupling, our
results on the induced AFM spin polarization of the NN Te atoms and the FM spin-polarization
of the next NN Sb atoms (see Table 1) are consistent with
the conclusion of Ref.~\onlinecite{JKim2017}. 

Fe-dopants are peculiar also when it comes to the possible source of magnetic order. As discussed above, the
Fe impurity states present in the bulk gap appear to form a band of localized impurity $d$-orbitals of minority-spin character. 
Note also that in the energy region of these states there is no presence of $p$-levels of the host material. On the basis of these results,
we are led to conclude that magnetic order in Fe-doped Sb$_2$Te$_3$ should be quite weak, if present at all.
Finally, Mn-doped Sb$_2$Te$_3$ closely resembles the dilute magnetic semiconductor (Ga,Mn)As, with a hole state
of $p$-character residing mostly on the NN atoms of the host, but hybridized with Mn $d$-levels.
Therefore Mn-doped Sb$_2$Te$_3$ should be characterized by a carried-mediated RKKY exchange coupling.
Note that, according to our calculations, Mn displays the largest magnetic moment of all four dopants. However, the presence of a large microscopic magnetic moment of the individual dopants does not guarantee the onset of
a robust magnetic order, which also depends on the dopant concentration and on the degree of $p$--$d$ hybridization.

\section{Samples growth}
Magnetically doped Sb$_2$Te$_3$ samples were grown by vertical Bridgman crystallization in conically shaped carbon-coated quartz ampoules. Elementary Sb, Te, V, Cr, Mn, and Fe were weighed to obtain the chemical composition Sb$_{2-x}$TM$_{x}$Te$_3$. After sealing under a residual pressure of $~1$ mbar, the ampoules were horizontally soaked at $700^\circ$ C for several days to ensure that magnetic impurities were homogeneously distributed. Subsequently, the samples were subjected to Bridgman recrystallization under an axial temperature gradient of $15^\circ$C/cm and ampoule translation rate of 10 mm/day. \cite{C3CE42026D}

\section{Scanning tunneling microscopy and spectroscopy}
\label{STM_STS}
Systematic STM/STS measurements have been performed with the goal of obtaining detailed information on the
impact that different doping elements have on both structural and electronic
properties of the host material. In this context, the high spatial
resolution of STM together with its capability of providing detailed
spectroscopic information for energies close to the Fermi level makes it an
ideal technique to visualize impurity-induced local changes in the density of
states emerging within the Sb$_2$Te$_3$ bulk band gap. 

Single crystals have been cleaved at room
temperature in ultra-high vacuum (UHV) and immediately inserted into the microscope
operated at a temperature $T = 4.8$\,K. All measurements have been performed
using electrochemically etched tungsten tips. Spectroscopic data have been
obtained using the lock-in technique by adding a bias voltage modulation in
between 1 and 10~meV (r.m.s.) at a frequency $f = 793$\,Hz.  Results for V,
Cr, Mn, and Fe are reported in Figs.~\ref{figV}, \ref{figCr}, \ref{figMn}, and \ref{figFe}, respectively.

\begin{figure}[t]   
	\begin{minipage}[t]{0.58\textwidth} 
		\includegraphics[width=\columnwidth]{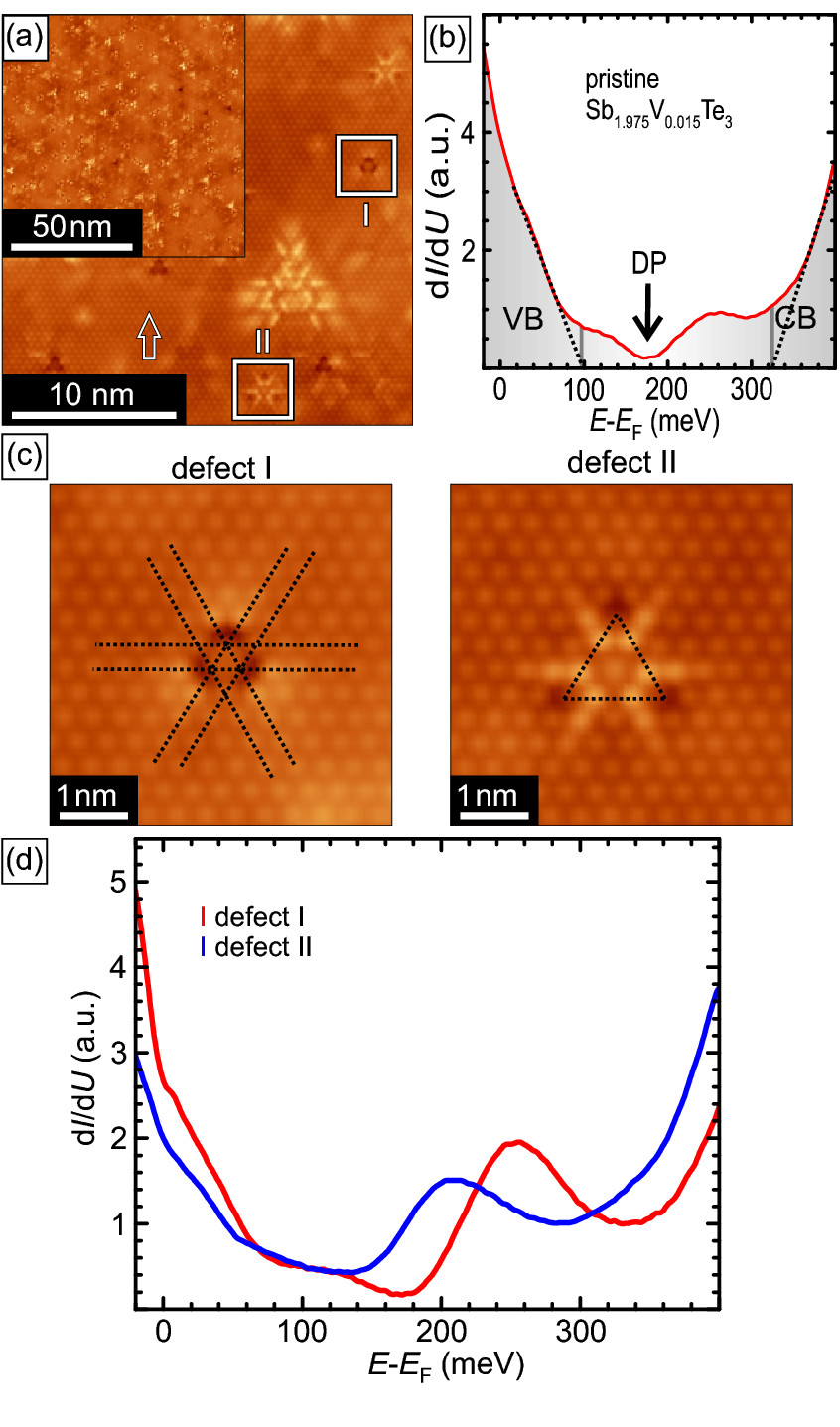} 
	\end{minipage}
	\hfill
	\begin{minipage}[b]{0.38\textwidth}	
		\caption{Scanning tunneling microscopy and spectroscopy data obtained on V-doped Sb$_2$Te$_3$. 
		(a) Topographic images reveal the existence of two different doping sites (see white boxes labeled I and II).  (b) Local density of states acquired over a defect-free area [see arrow in (a)]. (c) Atomically resolved images allow to precisely identify defects I and II as V atoms substituting Sb in the second and fourth layer of the Te-Sb-Te-Sb-Te quintuple layer structure, respectively. (d) Local density of states acquired by positioning the tip on top of the dopants.} \label{figV} \vspace{2cm}
	\end{minipage}	
\end{figure}     

\begin{figure}[t]   
	\begin{minipage}[t]{0.58\textwidth} 
		\includegraphics[width=\columnwidth]{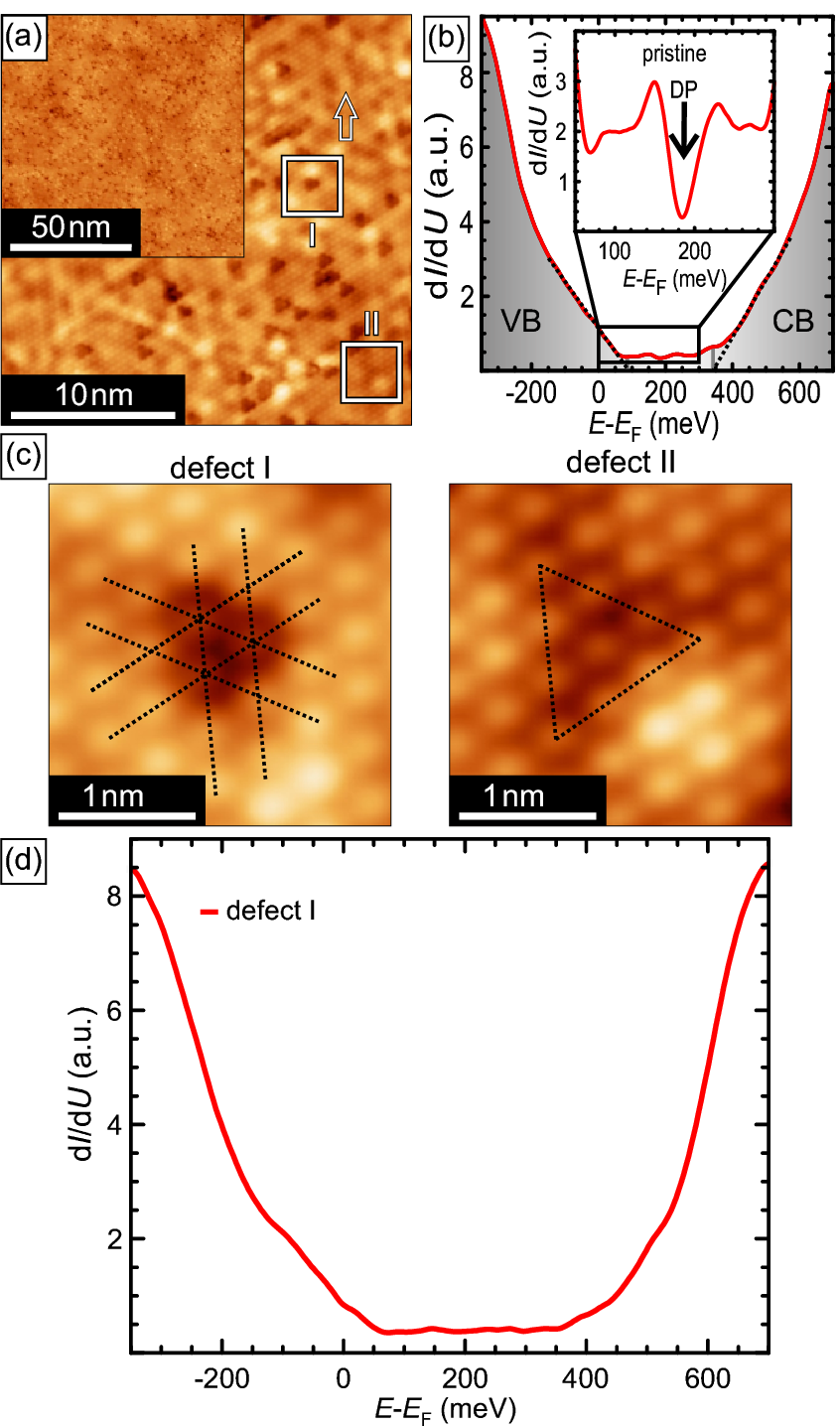} 
	\end{minipage}
	\hfill
	\begin{minipage}[b]{0.38\textwidth}	
		\caption{Scanning tunneling microscopy and spectroscopy data obtained on Cr-doped Sb$_2$Te$_3$. 
		(a) Topographic images reveal the existence of two different doping sites (see white boxes labeled I and II).  (b) Local density of states acquired over a defect-free area [see arrow in (a)]. (c) Atomically resolved images allow to precisely identify defects I and II as Cr atoms substituting Sb in the second and fourth layer of the Te-Sb-Te-Sb-Te quintuple layer structure, respectively. (d) Local density of states acquired by positioning the tip on top of defect I. The spectrum strongly resembles the one obtained over a defects-free area. In particular, not any new state emerges within the bulk-gap. The same holds for defect II. } \label{figCr} \vspace{2cm}
	\end{minipage}	
\end{figure}     

\begin{figure}[t]   
	\begin{minipage}[t]{0.58\textwidth} 
		\includegraphics[width=\columnwidth]{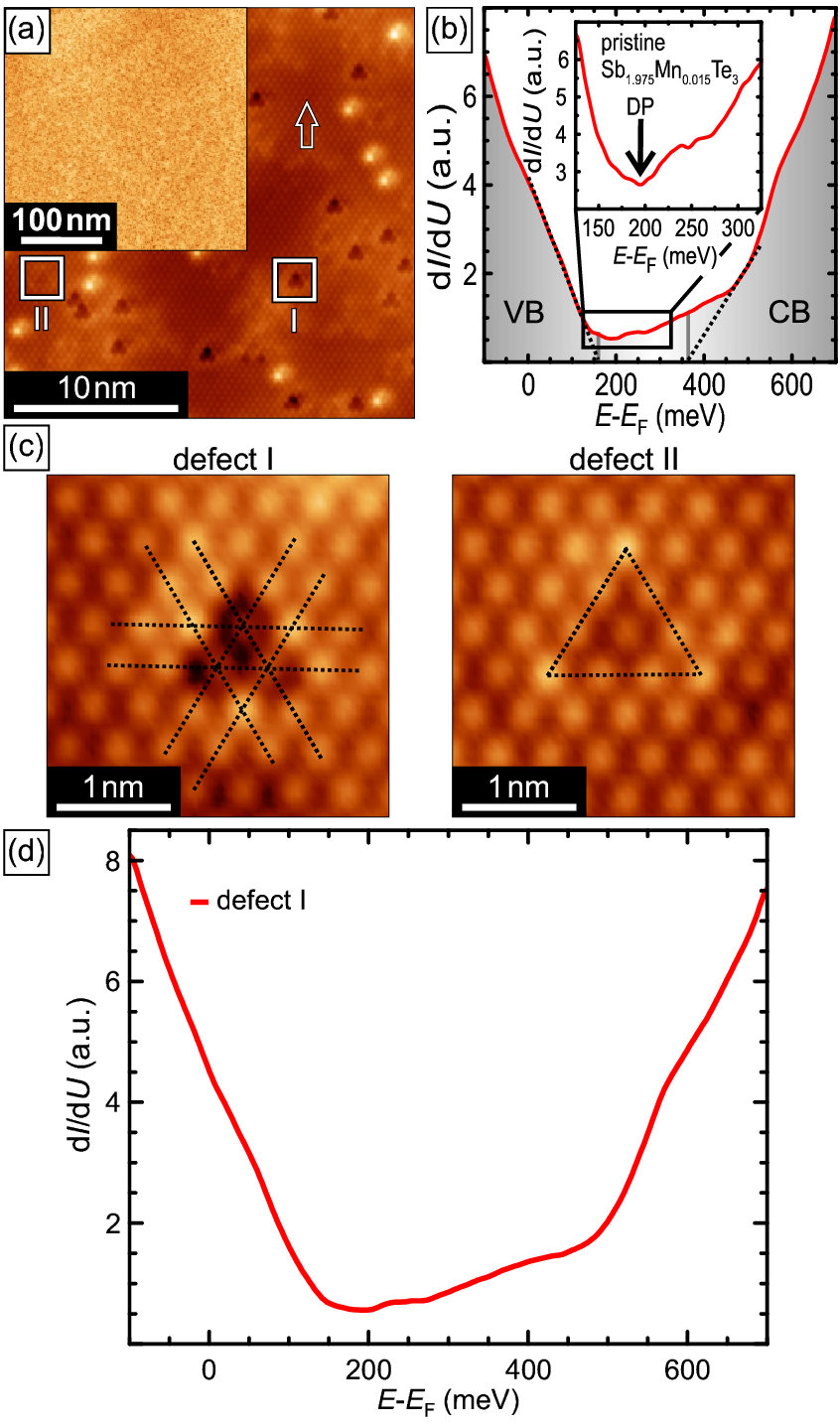} 
	\end{minipage}
	\hfill
	\begin{minipage}[b]{0.38\textwidth}	
		\caption{Scanning tunneling microscopy and spectroscopy data obtained on Mn-doped Sb$_2$Te$_3$. 
		(a) Topographic images reveal the existence of two different doping sites (see white boxes labeled I and II).  (b) Local density of states acquired over a defect-free area [see arrow in (a)]. (c) Atomically resolved images allow to precisely identify defects I and II as Mn atoms substituting Sb in the second and fourth layer of the Te-Sb-Te-Sb-Te quintuple layer structure, respectively. (d) Local density of states acquired by positioning the tip on top of defect I. The spectrum strongly resembles the one obtained over a defects-free area. In particular, not any new state emerges within the bulk-gap. The same holds for defect II.} \label{figMn} \vspace{2cm}
	\end{minipage}	
\end{figure}     

\begin{figure}[t]   
	\begin{minipage}[t]{0.58\textwidth} 
		\includegraphics[width=\columnwidth]{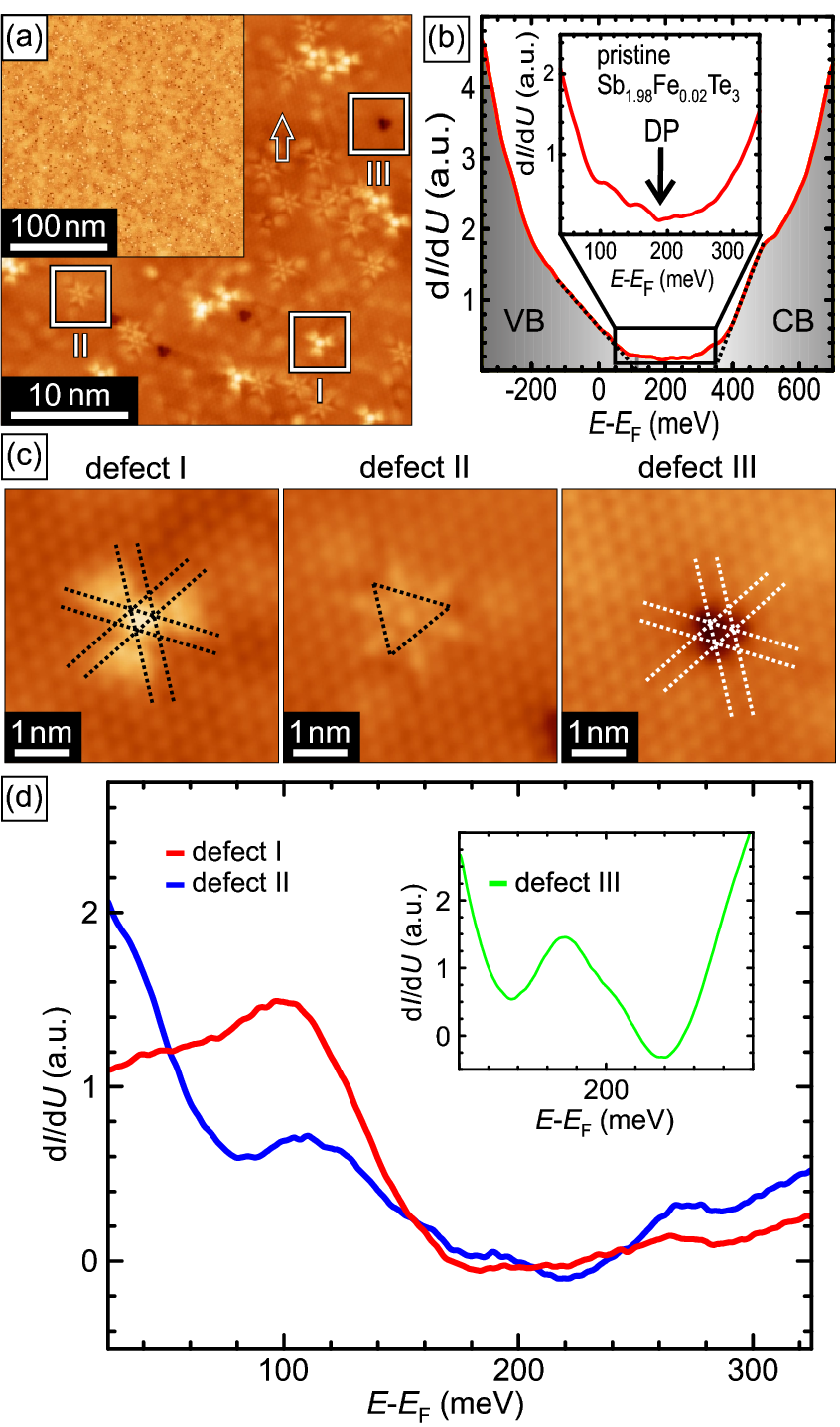} 
	\end{minipage}
	\hfill
	\begin{minipage}[b]{0.38\textwidth}	
		\caption{Scanning tunneling microscopy and spectroscopy data obtained on Fe-doped Sb$_2$Te$_3$. 
		(a) Topographic images reveal the existence of two predominant doping sites (see white boxes labeled I and II).  An additional but less frequently observed appearance is also found (see III). (b) Local density of states acquired over a defect-free area [see arrow in (a)]. (c) Atomically resolved images allow to precisely identify defects I and II as Fe atoms substituting Sb in the second and fourth layer of the Te-Sb-Te-Sb-Te quintuple layer structure, respectively. Defect III is
structurally equivalent to I but characterized by a different contrast. (d) Local density of states acquired by positioning the tip on top of the dopants.} \label{figFe} \vspace{2cm}
	\end{minipage}	
\end{figure}     

For all elements, topographic images reveal the presence of different defects in
the crystals. The large overviews reported in panels (a) as insets demonstrate that 
in all crystals the magnetic dopants are homogeneously
distributed, without any signature of clustering. This observation proves the
good quality of the samples used in our experiments, which can be classified as
dilutely doped.
Furthermore, irrespective of the particular TM dopant two types of defects,
labeled I and II, are predominant. These defects correspond to the
perturbations introduced in the crystal structure by the magnetic dopants, which
are expected to substitute Sb in the second and fourth layer of the
Te-Sb-Te-Sb-Te quintuple layer structure.\cite{sessi2016}  Although Sb substitution is energetically most favorable, 
interstitial doping has also been reported.\cite{watson2013} This necessarily calls for a precise identification
of the dopants sites. Indeed, the different local environment seen by impurities
residing at different sites can significantly change the impurity-host
hybridization, ultimately impacting onto the system's electronic and magnetic
properties.  As discussed in Ref.~\onlinecite{Jiang2012,Bathon2016}, this
information can be obtained by analyzing both the symmetry as well as the
spatial extension of the dopant-induced defects with respect to the atomically
resolved periodicity of the surface. As shown in panels (c), this method allows
us to identify defects I and II as magnetic atoms  substituting Sb in the second
and fourth layer of the Te-Sb-Te-Sb-Te quintuple layer structure, respectively.

The structural investigation already allows us to evidence some differences and
similarities among the different impurities. 
The cleavage planes of V- and Fe-doped samples have a quite similar appearance: 
dopants residing in the top-most Sb layer (defect I) are characterized by a triangular depression with brighter intensity at each corner. 
We would like to note that in the Fe case an additional but less frequently observed appearance is found
for dopants residing in the first Sb layer [defect III in Fig.\,\ref{figFe}(c)]. This site is
structurally equivalent to I but characterized by a different contrast. Although
a definitive conclusion can not be drawn at the moment, we speculate that defect
III might be related to Fe atoms in a 
different oxidation state. \cite{PhysRevB.88.235131} Dopants substituting the second Sb layer (defect II) appear as star-like features.
This scenario is different with respect to the one observed in Cr- and Mn-doped
samples, where triangular depressions are visible for both doping sites. Since
all magnetic TM elements are occupying the very same crystal site, they are always
perturbing the very same bonds. The different appearance visible on the surface
is thus a direct signature of a TM-dependent impurity--host hybridization. Along
this line, strong similarities are also expected to manifest in the electronic
structures, with V- and Fe-doped samples on one side and Cr- and Mn-doped samples on the other side.

Electronic aspects have been investigated by analyzing the local density of states as
inferred by STS measurements. Panels (b) show the results obtained by
positioning the tip away from any defects [see arrows in panels (a)]. As
described in Ref.\,\onlinecite{sessi2016,sessi22016}, the minimum visible in the
spectra marks the position of the Dirac point.  The sharp conductance increases 
above and below the Dirac point can be used to map the position of valence band maximum and
conduction band minimum. Their difference results in a bulk gap of approximately
175\,meV, in good agreement with previous reports.\cite{Jiang2012,sessi2016,sessi22016}  

The impact of the dopants onto the sample electronic properties has been analyzed 
by acquiring STS spectra with the tip positioned directly
on top of the two different magnetic dopants (I and II).
For both V and Fe, these spectra [reported in panels (d)] highlight the emergence of
broad resonances residing within the bulk gap at energies close to the Dirac point. 
Quite contrary, data acquired over Cr and Mn dopants do not evidence
any additional feature, as these spectra are virtually identical to those obtained in defects-free samples areas.  
These experimental observations are in agreement
with the theoretical calculations presented in Sec.\,\ref{Subsec:ImpRes} which showed that V and Fe atoms 
generate a substantial impurity-induced spectral weight at the Dirac point, whereas Cr and Mn
leave the Dirac node almost unaffected (see Fig.~\ref{SurfTeDOS}). Overall,
these findings indicate that magnetic impurities can profoundly alter the
density of states within the bulk gap, making the observation of a
magnetic-induced opening in the surface Dirac cone impossible. This is not only
the case in V- and Fe-doped samples, where strong impurity resonances can
effectively fill the magnetic gap. Indeed, although being strongly suppressed, a
non-vanishing contribution to the total density of states is present also for Cr
and Mn. This is clearly evidenced by the Cr-doped sample where, despite the
presence of long-range ferromagnetic order (see XMCD data and related
discussion), the surface remains gapless as highlighted by the zoomed STS
spectrum [see Fig.\,\ref{figCr}(b): the differential conductivity never goes to zero but simply reaches a minimum].

\section{Resonant Photoemission Measurements}   
\label{resPES}

\begin{figure}[t]   
	\begin{minipage}[t]{0.58\textwidth} 
		\includegraphics[width=\columnwidth]{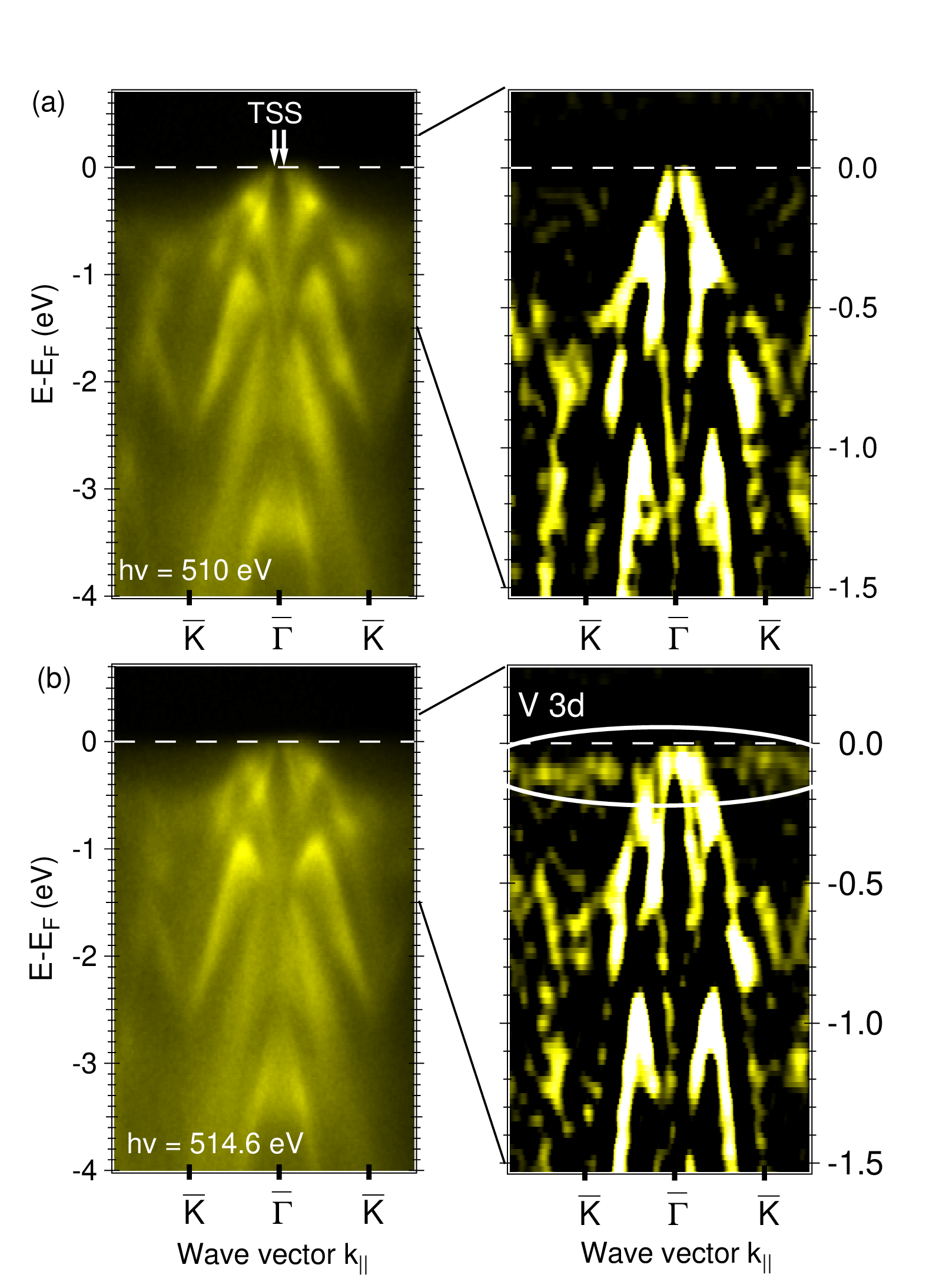} 
	\end{minipage}
	\hfill
	\begin{minipage}[b]{0.38\textwidth}	
		\caption{Angle-resolved resonant photoemission data for V-doped Sb$_2$Te$_3$. 
		On- and off-resonant data sets, as well as corresponding 2$^{nd}$-derivative images, are shown in (a) and (b),
		respectively (cf.\ Fig.~\ref{fig10}(a) for the corresponding $L_3$ XAS spectrum). 
		Despite the strong $p$-doping of Sb$_2$Te$_3$, the tails of the topological surface state (TSS) 
		extending into the valence band are observed.\cite{Seibel:15} 
		In the on-resonant data in (b) a dispersion-less feature is observed near the Fermi level, 
		which is attributed to a V 3$d$ impurity band.\cite{peixoto2016}  
		} \label{fig9} \vspace{2cm}
	\end{minipage}	
\end{figure}     

In this section we shall discuss resonant photoemission data that were obtained for the same 
V-, Cr-, Mn- and Fe-doped Sb$_2$Te$_3$ samples studied by STM/STS in the previous section. 
This method provides a rather direct estimate of the 3$d$
impurity DOS in the valence band by exploiting their strongly
enhanced photoemission cross section for excitation energies close to the
2$p$--3$d$ x-ray absorption (XAS) maximum.\cite{Mulazzi:10,Kobayashi:14} 
Typically this is achieved by considering the difference between an
``on-resonant'' spectrum, obtained at the XAS maximum, and an ``off-resonant''
spectrum, obtained at an energy just below the XAS threshold. The results can be
directly compared to the calculated DOS in section II, thus complementing the STS data which
cannot access electronic states residing well below the Fermi level.

The resonant photoemission data were obtained at beamline P04 of the PETRA III storage ring at DESY (Germany) using the
ASPHERE III setup designed for soft X-ray photoemission experiments. The measurements were performed with a Scienta R4000 spectrometer at a base pressure of ca. 3$\times$10$^{-10}$~mbar and a temperature of ca. 30~K. The energy resolution was ca. 100~meV and for all measurements circularly polarized light has been used.  

Figures~\ref{fig9}(a) and (b) display angle-resolved off- and on-resonant
photoemission data for V-doped Sb$_2$Te$_3$ showing well-defined band dispersions. 
Due to the strong intrinsic $p$-type doping the Dirac point of the topological surface state (TSS) 
lies above the Fermi level and, hence, is inaccessible to photoemission. 
This is in agreement with our STS results in Fig.~\ref{BulkDOS}. 
However, the high-binding-energy tails of the TSS extend
into the valence band up to a binding energy of about 0.3~eV, as shown
previously for undoped Sb$_2$Te$_3$ by high-resolution ARPES.\cite{Seibel:15}
Remarkably, the TSS can also be discerned in the present soft X-ray ARPES data, as indicated in Fig.~\ref{fig9}(a).  

\begin{figure*}[t] \includegraphics[width=\columnwidth]{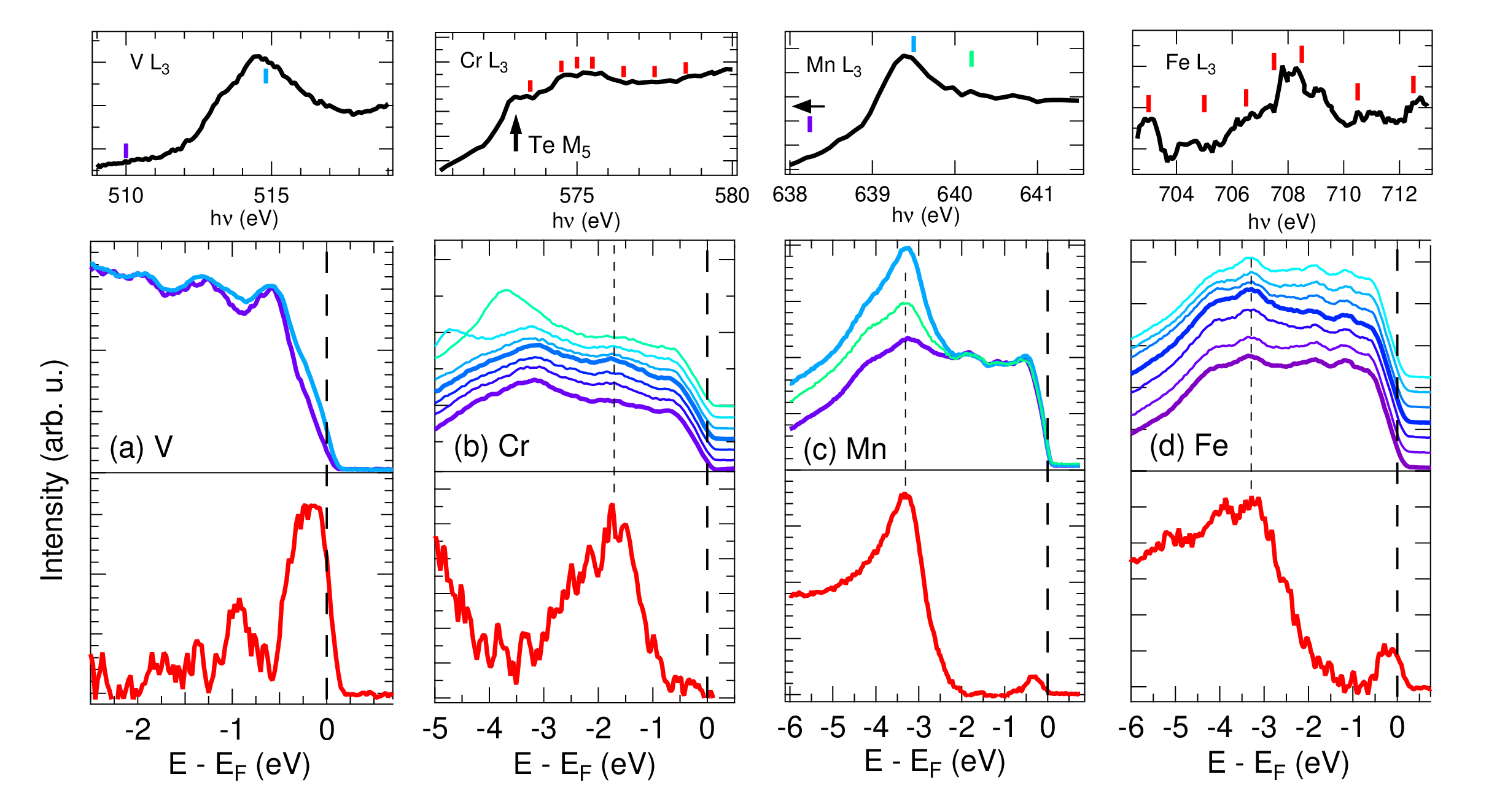}  
\caption{Resonant photoemission data for (a) V-, (b) Cr-, (c) Mn-, and (d) Fe-doped Sb$_2$Te$_3$.
The top panels show the respective 2$p$-3$d$ $L_3$ absorption edges. The markers indicate the photon energies $h\nu$ used for the resonant photoemission measurements shown in the middle panels. The photoemission spectra in (b) and (d) are offset, with the bottom (top) spectrum corresponding to the lowest (highest) photon energy. The off-resonant spectrum in (c) was acquired at $h\nu =$~635~eV, as indicated by the arrow. The pronounced features at higher binding energy in the two topmost spectra in (b) are artifacts arising from the second-harmonic of the beamline and the energetic overlap between the Cr $2p$ and Te $3d$ core levels (see XAS spectrum). The bottom panels show the difference between an on-resonant and an off-resonant spectrum, which are indicated by thick lines in the middle panels. The difference spectra provide an estimate of the 3$d$ impurity DOS that can be compared to the corresponding theoretical calculations in Fig.~\ref{BulkDOS}.     
} \label{fig10} \end{figure*}  

In order to address the V 3$d$ states we consider the 2$^{nd}$ derivative images in
Figs.~\ref{fig9}(a) and (b) [right panels], that were obtained from the raw-data sets (left panels). 
In the on-resonant image one can discern a weak, dispersion-less feature near the Fermi level 
that is absent in the off-resonant image and, hence, is attributed to a V 3$d$ impurity band. 
The presence of V states near the Fermi level is further demonstrated by the angle-integrated data in Fig.~\ref{fig10}(a). 
The on-resonant spectrum shows a shoulder at a binding energy of ca.~0.15~eV, 
leading to a line-shape deviation from the off-resonant spectrum.
The resulting difference spectrum indicates a metallic V 3$d$ DOS extending down
to a binding energy of ca.~0.5~eV. This observation is in good agreement with
our theoretical calculation of the V DOS shown in Fig.~\ref{BulkDOS} and with previous measurements for V-doped (BiSb)$_2$Te$_3$ thin films \cite{peixoto2016}. 
In particular, the direct comparison appears to favor the calculation without an
additional on-site interaction parameter $U$, possibly indicating a minor role of
correlation effects. Interestingly, our photoemission data demonstrate an energy
overlap of V impurity states and the TSS, facilitating an interaction between
them, as observed in our STS experiments in Sect.\,\ref{STM_STS}
and in previous reports.\cite{sessi2016} 

Resonant photoemission data for Cr-doped Sb$_2$Te$_3$ are displayed in Fig.~\ref{fig10}(b). 
The photon energy-dependent spectra across the Cr $L_3$-edge
show a resonating feature at a binding energy of ca.~1.7~eV. 
Accordingly, the difference spectrum indicates a maximum in the Cr 3$d$ DOS at this energy. 
This finding is in good agreement with our theoretical results for Cr reported in Fig.~\ref{BulkDOS}.

Figure~\ref{fig10}(c) shows resonant photoemission data for Mn-doped Sb$_2$Te$_3$. 
We find a pronounced Mn 3$d$ feature at a binding energy of ca.~3.3~eV. Furthermore, an
additional hump centered at ca.~0.3~eV binding energy leads
to a finite DOS at the Fermi level. These observations are in good agreement with our
calculations (see Fig.~\ref{BulkDOS}). Similar features in the Mn DOS have also been found for (Ga,Mn)As.\cite{Kobayashi:14}  

The resonant photoemission data for Fe-doped Sb$_2$Te$_3$ reported in Fig.~\ref{fig10}(d) 
show a main peak centered at a binding energy slightly below 3.0~eV. This is in line with our theoretical predictions which evidence a maximum in the Fe 3$d$ majority states at similar energies (see Fig.~\ref{BulkDOS}). The experimental data also suggest the presence of Fe states near the Fermi level, which our calculations can directly link to the presence of a metallic 3$d$ minority band. 

\section{XMCD Measurements}
\label{xmcd}

To probe the magnetic properties of the TM dopants
embedded in the TI host, XMCD measurements were performed at the BOREAS beamline
of the ALBA synchrotron light source\cite{Barla2016} and at the ID32 beamline of the European Synchrotron Radiation Facility. \cite{Kummer2016} The samples were cleaved
\textit{in situ} under UHV conditions. The experiments were carried out at
 temperatures $T < 10$\,K in the total electron yield mode,
which makes our measurements mostly surface sensitive. In Fig.~\ref{XMCD} we
plot the $L$ edge absorption spectra for left and right circularly polarized
light (red and blue curves, respectively) and their corresponding XMCD spectra
(black curves) for each of the Sb$_2$Te$_3$ system doped with (a) V, (b) Cr, (c)
Mn, and (d) Fe, measured under a magnetic field applied along
the surface normal. The strong contribution of the background in the XAS
line shapes is related to the low concentration of the dopants. The Sb$_2$Te$_3$
single crystals exhibited slightly different dopant concentrations, namely 1.5\%
V, 1.5\% Cr, 1.5\% Mn and 0.5\% Fe in each Sb layer. 
Despite of the complex multiplet fine structure of the TM $L$-edges and the clear overlap between the Cr $L$- and Te
$M$-edges, no sign of contaminated (oxidized) TM species were observed in the
system. A clear XMCD signal is detected in all our samples, proving that V, Cr, Mn, and Fe dopants carry a net magnetic moment once embedded into Sb$_2$Te$_3$. Our results for V- and
Cr-doped Sb$_2$Te$_3$ qualitatively reproduce previous XMCD studies,\cite{sessi2016,Ye2015} 
whereas no similar study has been found for the Mn- and
Fe-doped systems.

\begin{figure*}[t] \centering \includegraphics[width=\columnwidth]{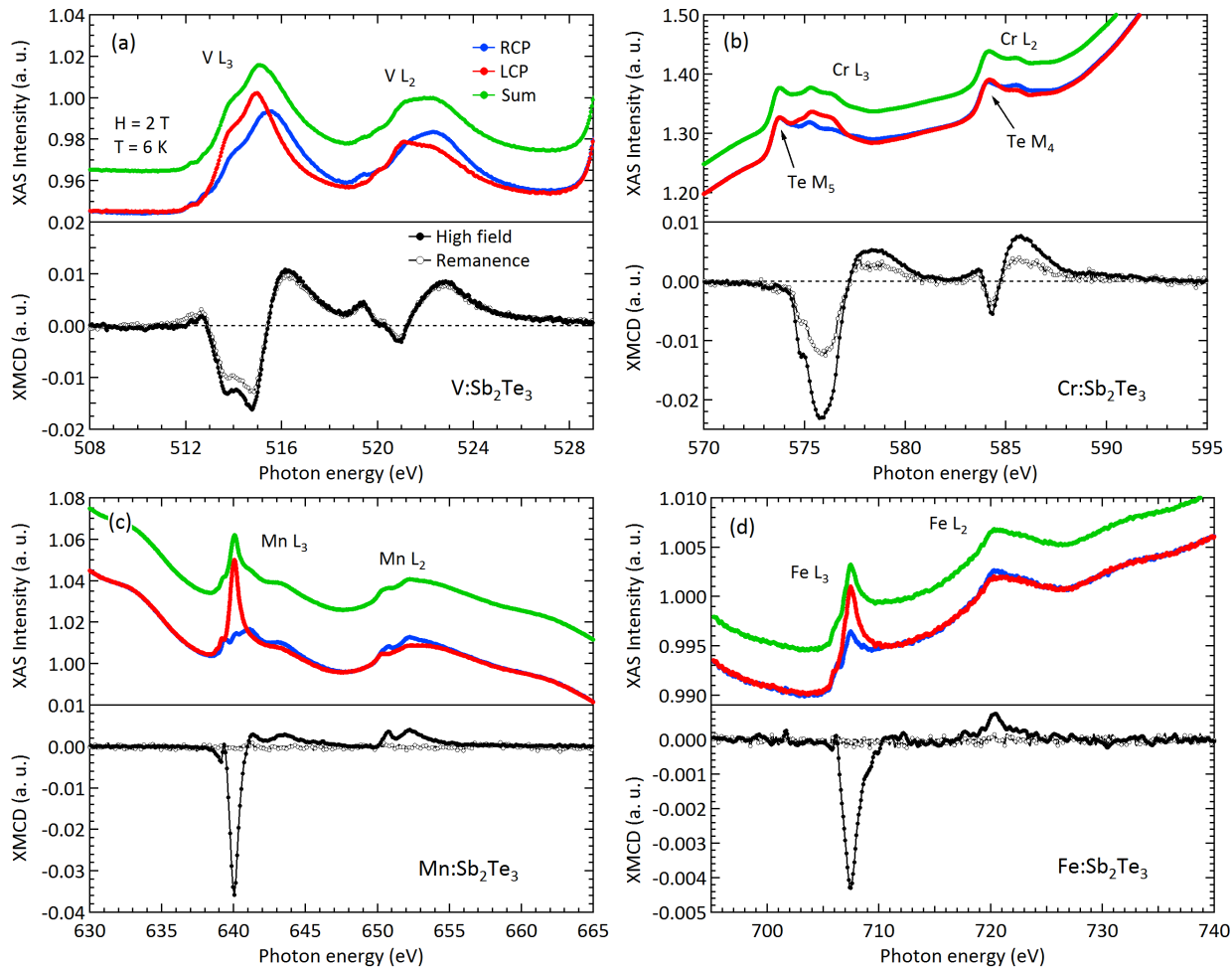} 
\caption{X-ray absorption (XAS) and magnetic dichroism (XMCD) spectra at the
$L_{2,3}$ edges ($2p-3d$ transition) of the TM element from (a) V-, (b)
Cr-, (c) Mn-, and (d) Fe-doped Sb$_2$Te$_3$ single-crystals. The XAS spectra for left (LCP) and
right (RCP) circularly polarized light are shown in the upper panels (red and
blue curves, respectively). The corresponding XMCD spectra, defined as
$(R-L)/(R+L)$, are shown as the black curves in the lower panels, and reveal a
sizable localized magnetic moment carried by the $3d$ states.} 
\label{XMCD}
\end{figure*}   

Although doping with V, Cr, Mn, and Fe always results in the introduction of magnetic moments into the system, the long-range magnetic properties can be strongly TM-dependent. This is expected because of the element-specific $p$--$d$ hybridization between dopant and host material, which plays a crucial role in mediating the coupling between magnetic moments in TM-doped TIs.

\begin{figure}[b]   
	\includegraphics[width=0.6\columnwidth]{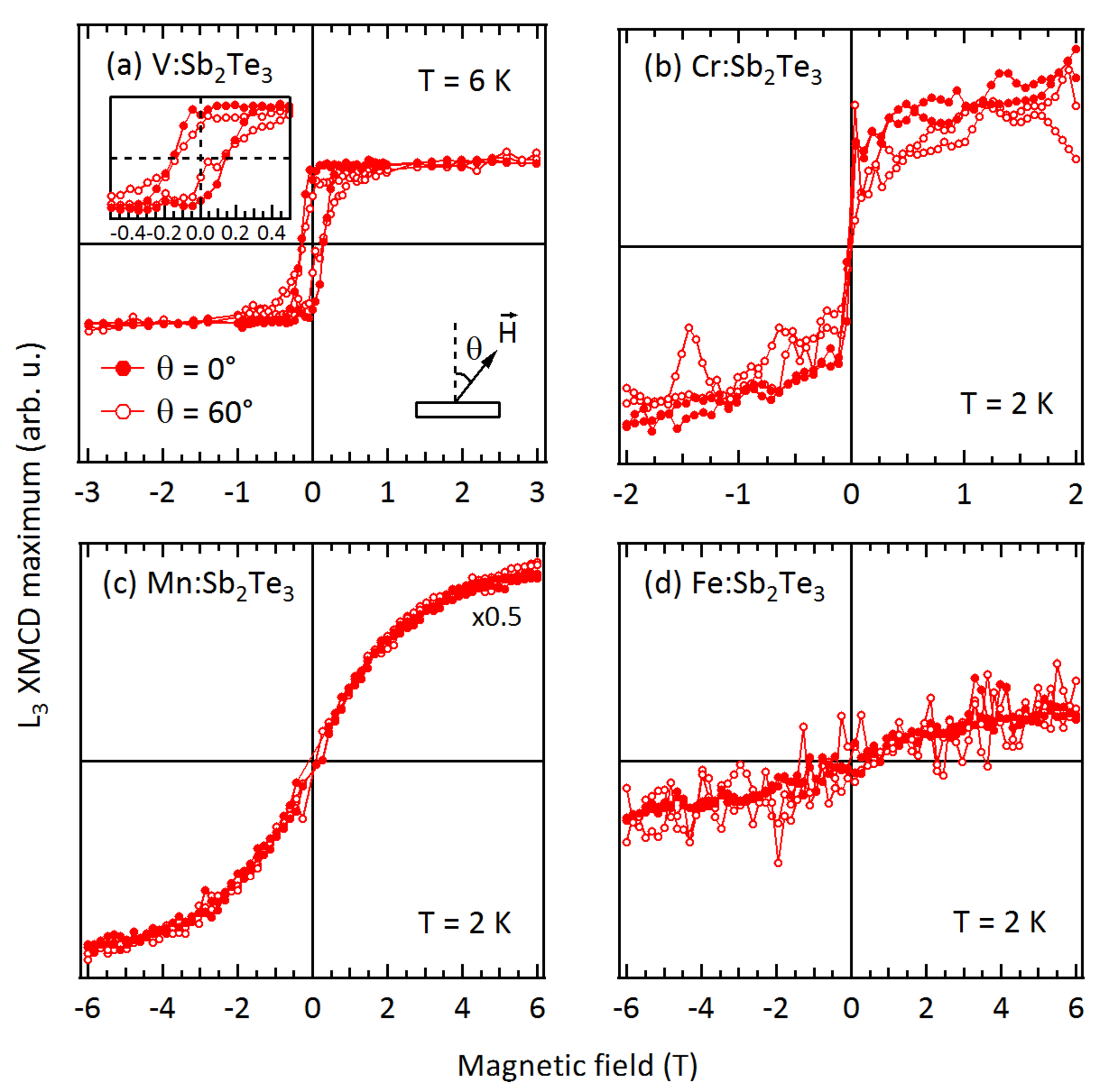}  	
		\caption{Hysteresis loops taken at the $L_3$ XMCD maximum of each TM element 
		for (a) V-, (b) Cr-, (c) Mn-, and (d) Fe-doped Sb$_2$Te$_3$ single-crystals  
		for magnetic field applied at angles $\theta = 0^\circ$ (full circles) 
		and $\theta = 60^\circ$ (hollow circles) with respect to the surface normal. 
		V- and Fe-doped samples show a paramagnetic response at these low concentrations (see text), 
		whereas the Mn-doped samples are rather isotropic and Cr-doped samples 
		tend to favor an out-of-plane easy axis, as expected for QAH systems. 
		The different macroscopic magnetic properties are determined by the doping concentration 
		and the strength of the $p$--$d$ hybridization between the impurity and host states, 
		which ultimately determine the dominant magnetic interactions in the system.} \label{Hyst} 
\end{figure}     

These aspects have been investigated by measuring hysteresis loops at the $L_3$ XMCD maximum, for a
magnetic field applied at different angles $\theta$ with respect to the surface
normal. The systems studied here showed very different macroscopic magnetic
behaviors. In Fig.\,\ref{Hyst} we show the resulting magnetization curves for
$\theta = 0^\circ$ (full circles) and $\theta = 60^\circ$ (hollow circles) for (a)
V-, (b) Cr-, (c) Mn-, and (d) Fe-doped systems. For the V- and Cr-doped samples,
a ferromagnetic hysteresis curve can be seen, showing a clear trend
towards an out-of-plane ground-state magnetization. 
In particular, the V-doped sample exhibits an opening of the hysteresis loop, with a
coercivity of approximately 150\,mT, as seen in the inset in Fig.\,\ref{Hyst}(a).
For Cr, a collective ferromagnetic response is highlighted by the step-like
behavior of the hysteresis curve. However, the coercivity in this case could
not be measured, being it smaller than the resolution of our measurements. The
Mn-doped sample showed the largest XMCD signal, which is indicative of a high
spin configuration of the dopants. However, long-range magnetic order is absent
in this system: the hysteresis loops show no sign of saturation up to a field
of 6\,T. Furthermore, data taken along normal and grazing magnetic field
directions are indistinguishable, indicating the absence of any detectable preferred
magnetization direction. Finally, the Fe-doped sample showed a 
paramagnetic behavior. The low concentration results in a small signal which makes it difficult to draw definitive conclusions over the magnetic easy-axis. 

As previously discussed, the magnetic impurities are expected to induce a magnetic moment
in the neighboring Te and Sb atoms via $p$--$d$ hybridization. This has been
directly detected by XMCD measurements. Fig.\,\ref{SbM45} (upmost panel) shows
an example of the Sb $M_{4,5}$ edge line shapes for left and right
circularly polarized light (blue and red curves, respectively), measured at a
Cr-doped sample. They show no sign of oxygen contamination (O K-edge lies at 535
eV) and are highly reproducible in all samples. The lower panels show the Sb
$M_{4,5}$ XMCD spectra for samples doped with the different transition metals,
as labeled in the figure. The intensity of this signal
depends on the dopant concentration and the degree of $p$--$d$
hybridization. In V, Cr, and Mn-doped samples, our measurements reveal a small magnetic
dichroism at the Sb $M_{4,5}$ edge.  The XMCD signal has an
opposite sign as compared to the corresponding TM $L$ edges. This proves the existence of ferromagnetic coupling
between TM and Sb ions. All these observations are in agreement with our theoretical predictions. The Fe case could not be experimentally scrutinized. The magnetic polarization of Sb induced by Fe, if any, is below our detection limit due to the three times lower amount of Fe in the sample, as compared to the other 3d elements.

\begin{figure}[t] \centering \includegraphics[width=3in]{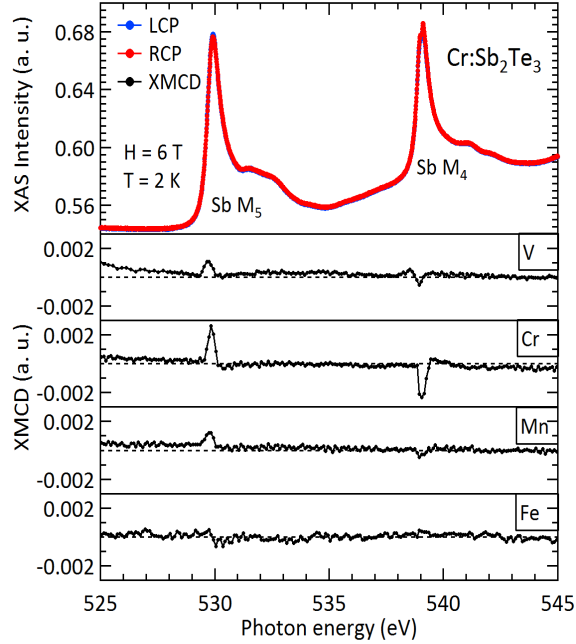}
\caption{XAS and XMCD spectra at the Sb $M_{4,5}$ edges ($3d-5p$ transition). Upmost panel: XAS line shape for incident left (blue) and
right (red) circularly polarized light, from a Cr-doped Sb$_2$Te$_3$ sample. 
Lower panels:  corresponding XMCD spectra (black curves). 
The XMCD signal seen at the Sb $M$ edges
is attributed to a partial spin-polarization of the Sb $p$ states at
the valence band via $p$--$d$ hybridization with the impurity states of the
neighboring TM ions.}
\label{SbM45} \end{figure}

Finally, we would like to compare the experimental and theoretical magnetic anisotropy directions. For Cr-doped Sb$_2$Te$_3$, our calculations predict an out-of-plane easy axis for both surface and bulk dopants, perfectly in line with the experimental results.  On the other hand, for V-doped samples our calculations
predict a strong in-plane anisotropy at the surface and a weak out-of-plane anisotropy in the bulk. This seemingly apparent discrepancy between theory and experiments can be explained by considering the XAS/XMCD probing depth. Indeed, although being very surface sensitive, our measurements do not exclusively access the very top layer but probe the sample up to a thickness of a few nanometers. Furthermore, the presence of bulk long-range ferromagnetic order can effectively act onto the surface magnetic moments, forcing them to align along the surface normal.
This observation also explains the experimental data obtained on Mn-doped samples. According to our calculations, Mn surface and bulk impurities are predicted to have in-plane and out-of-plane easy axis, respectively, with the strength of the magnetic anisotropy energy being roughly the same for the two cases. Summing these two contributions results in the absence of any preferred direction, in agreement with our experimental evidence. In conclusion, we can say the experimental results for the magnetic anisotropy axis are consistent with the results of the theoretical calculations for those cases, like Cr, where theory predicts the same easy axis for both bulk and surface doping. For the impurities where this is not the case, such as V, the dominant contribution is the result of a subtle interplay between in-plane and out plane anisotropies. For these cases the final outcome might depend on details difficult to be implemented theoretically.

\section{Conclusions}
\label{conclusions}

In this work we have investigated the electronic and magnetic properties of
Sb$_2$Te$_3$ doped with different substitutional transition
metal impurities (namely V, Cr, Mn, Fe). Our calculations show that magnetic
impurities induce spin polarization at the host atoms. Both in bulk and in
surface doping V, Cr and Mn couple anti-ferromagnetically with the NN Te atoms,
while Fe couples ferromagnetically. In all cases the impurity atom couples
ferromagnetically with NN Sb host atoms. 

The magnetic anisotropy of TM-doped Sb$_2$Te$_3$ depends not only on the type of
TM impurity but also on its position. Specifically, the theoretical results have shown that while
the easy axis for V and Mn is out-of-plane for bulk impurities, 
it becomes in-plane for surface impurities.  For Cr doping the easy
axis is along the out-of-plane direction for both bulk and surface doping.  
Fe is the only impurity displaying an in-plane anisotropy for both bulk and surface doping.

One of the important conclusions of this work is that substitutional magnetic TM
impurities not necessarily open a gap at the DP. This apparent contradiction is effectively explained 
by the possible presence of impurity resonances. Specifically, in the case of V and Fe 
we find that impurity states appear to significantly contribute to the density of states at energies close to the DP. 

These theoretical predictions have been compared with experimental results. 
In agreement with DFT calculations, STS data revealed that both V and Fe dopants introduce 
strong impurity states within the bulk gap, whereas no resonances are detected for Cr and Mn. 
The use of resPES allowed us to directly access the 3$d$ character of the impurity states, 
further corroborating the validity of our calculations.  
Finally, the magnetic properties have been experimentally scrutinized by XMCD measurements. 
All dopants have been found to induce a magnetic moment in the Sb atoms, signaling the $p$--$d$ 
hybridization taking place between impurities and host material. On the other hand, 
long-range magnetic order is established only for V and Cr doped samples.

The results of the present joint theoretical and experimental study have implications for the understanding of the microscopic
nature of the QAHE in magnetically doped TIs, and provide
indications of the most promising systems that could display enhanced forms of QAHE.
One conclusion of our work is that, for the same host material Sb$_2$Te$_3$,
the compliance of the canonical conditions 
for the realization of the QAHE -- robust long-range ferromagnetic order with out-of-plane anisotropy 
and absence of bulk conducting
states in the vicinity of the Dirac point -- depends crucially on the nature of the doping. For this reason, Mn and Fe are not
expected to satisfy the conditions for the onset of the QAHE, mainly due to the lack of a robust ferromagnetic order.
Of the other two impurities, Cr is the most clear-cut system to host a robust QAHE: it displays a reasonably strong out-of-plane
ferromagnetic order (in this case our theoretical results agree with experiment) 
and no extra states appear in the bulk gap as a result of doping. Our results for Cr are in agreement with a recent theoretical
study,\cite{JKim2017} which for these reasons finds Cr-doped Sb$_2$Te$_3$ (and BiSb$_2$Te$_3$) to be the best candidate to host a robust QAHE
and suggests ways to further
enhance the temperature range where it can occur.   
V doping is perhaps the most interesting
and challenging case of the four investigated. For this impurity, XMCD results indicate the presence of 
strong ferromagnetic order with an 
out-of-plane easy-axis. 
On the other hand, the presence of impurity states in the bulk gap of TI in the vicinity of Dirac point is now confirmed by both
experimental and theoretical studies and defies the second condition for the QAHE.\cite{sessi2016} 
The observation of a clear QAHE for this system\cite{chang2015}, essentially
with the same degree of precision found in  Cr-doped Sb$_2$Te$_3$, 
can possibly be explained by invoking the existence of a mobility gap,
similar to the one responsible for the perfect Hall quantization in GaAlAs 2D electron gas systems. Given this more complex
scenario, further rational quantum engineering and material optimization of V-doped Sb$_2$Te$_3$ TI, for example by co-doping, in order
to enhance its QAHE properties, is probably more challenging than for Cr-doped TIs.


\section{Acknowledgments} 

This work was supported by the Faculty of Technology and by the Department of
Physics and Electrical Engineering at Linnaeus University (Sweden).  C.M.C and
F.I acknowledge financial support from the Swedish Research Council (VR) through
Grant No.  621-2014-4785, and by the Carl Tryggers Stiftelse through Grant No.
CTS 14:178.  Computational resources have been provided by the Lunarc Center for
Scientific and Technical Computing at Lund University. The XAS and XMCD
experiments were performed at the BOREAS beamline at ALBA Synchrotron and at the ID32 beamline of the European Synchrotron Radiation Facility. O.E.T. and K.A.K. have been supported by the Russian Science Foundation (project No.17-12-01047). This work was supported by the DFG through SFB1170
'ToCoTronics' (projects A01and A02).

\bibliography{SIA_cmc2017}

\end{document}